\documentclass[reprint,showpacs,preprintnumbers,superscriptaddress,multibbl,amsmath,amssymb,prl]{revtex4-1}
\usepackage{graphicx,wasysym}
\usepackage{dcolumn}
\usepackage{bm}
\usepackage{tabulary}
\usepackage{color}

\newcommand{\vo}{v_\mathrm{o}}

\usepackage{dsfont}

\begin{document}

\title{Current fluctuations of interacting active Brownian particles}

\author{Trevor GrandPre}
\affiliation{%
Department of Physics, University of California, Berkeley 
}
\author{David T. Limmer} \email{dlimmer@berkeley.edu}
\affiliation{%
Department of Chemistry, University of California, Berkeley 
}
\affiliation{%
Kavli Energy NanoSciences Institute, University of California, Berkeley 
}\affiliation{%
Lawrence Berkeley National Laboratory, University of California, Berkeley 
}
\date{\today}

\begin{abstract}
We derive the distribution of particle currents for a system of interacting active Brownian particles
in the long time limit using large deviation theory and a weighted many body expansion. We find the
distribution is non-Gaussian, except in the limit of passive particles. The non-Gaussian fluctuations
can be understood from the effective potential the particles experience when conditioned on a given
current. This potential suppresses fluctuations of the particles orientations and surrounding density,
aligning particles and reducing their effective drag. From the distribution of currents, we compute
the diffusion coefficient, which is in excellent agreement with molecular dynamics simulations over a
range of self-propulsion velocities and densities. We show that mass transport is Fickian in that the
diffusion constant determines the response of a small density gradient, and that nonlinear responses
are similarly computable from the density dependence of the current distribution.

\end{abstract}

\maketitle
Persistent currents are the hallmark of a system driven away from equilibrium. 
One of the simplest and most fundamental problems of nonequilibrium physics is to predict the structure of the fluctuations of currents around a nonequilibrium steady-state and  to decode the microscopic information contained in them. Non-equilibrium fluctuation-dissipation relations~\cite{Speck2006,Prost2009,Baiesi2009,Seifert2010,Chetrite2011,Baiesi2013,Maes2014}, fluctuation theorems~\cite{Evans1994,Kurchan1997,Crooks1999,Seifert2005,Jarzynski1997,Gallavotti1995,Lebowitz1998b}, and thermodynamic uncertainty relations~\cite{Barato2015,Gingrich2016,Polettini2016} are notable examples of successes towards this end. Much of this progress has been underpinned by the study of large deviation functions (LDFs), which supplies a general framework to compute and characterize fluctuations of extensive observables~\cite{Chetrite2015,Lebowitz1998}. The LDFs of the current can be viewed as the analog of a free energy, making relationships between fluctuations and response to external perturbations transparent~\cite{Speck2016,Gao2017,Gaspard2017}. However, the evaluation of LDFs for interacting systems remains challenging. In this paper, we characterize the fluctuations of currents in a system of interacting active Brownian particles (ABPs) and show how these fluctuations encode the response of the system.

ABPs are a simple model of active matter, a class of systems that convert energy from the environment into directed motion. ABPs evolve nonequilibrium steady states as they break detailed balance at the single particle level due to a constant nonconservative driving force. More than just being non-Boltzmann, their steady-states support unique phenomena such as motility induced phase separation~\cite{Cates2015,Redner2013}. Laboratory realizations of active matter include cellular biopolymers~\cite{Schaller2010,Sumino2012,Schaller2011},  bacteria~\cite{Berg1972,Shapiro1995,Dombrowski2004,Hill2007,Lauga2006,Fu2012,Parsek2005}, and synthetic colloids~\cite{Palacci2013,Narayan2006,Howse2007,Walther2008,Bricard2013}, with the latter being a direct realization of ABPs\cite{Bechinger2016,Buttinoni2013,Buttinoni2013,Elgeti2015}. Indeed it has been demonstrated that the center of mass motion for bacteria and biopolymers can be well described by ABPs with an effective particle size when hydrodynamic interactions and internal degrees can be neglected\cite{Yang2010,Peruani2016,Ginelli2010,Abkenar2013,Peruani2012,Harvey2013,Chelakkot2014,Elgeti2015}. 

We derive the current LDFs for ABPs and validate it with molecular simulation. We find that small current fluctuations are Gaussian, and the associated linear response obeys Fick's law, as has been shown for noninteracting ABPs~\cite{wagner2017steady}. Large current fluctuations are non-Gaussian and the associated nonlinear response results from a change in the particle's orientational correlations, which we characterize with the effective potential that renders those fluctuations typical.

We consider a collection of $N$ ABPs in two spatial dimensions, whose positions and orientations are denoted $\textbf{r}^N=\{ \textbf{r}_1,\textbf{r}_2 \dots \textbf{r}_N \}$ and $\boldsymbol{\theta}^N= \{ \theta_1,\theta_2 \dots \theta_N \}$, respectively. These dynamical variables are coupled through their equations of motion, which for the position of the $i$th ABP  is
\begin{equation}
\dot{\textbf{r}}_i(t) =\textbf{F}_i\left [\textbf{r}^N(t)\right ]+{\vo}\textbf{e} \left [ {\theta_i}\left(t\right)\right ]+\boldsymbol{\eta}_{\mathrm{t}}\left(t\right),
\end{equation}
and for it's corresponding orientation is
\begin{equation}
\dot{\theta}_i(t) ={\eta_{\mathrm{r}}}\left(t\right) \, ,
\end{equation}
where the dot denotes time derivative, $\vo$ is the magnitude of the self-propulsion velocity, $\boldsymbol{e}\left [ {\theta}\left(t\right)\right ]=\{ \cos(\theta), \sin(\theta) \}$ is the unit vector on a circle.  The Gaussian random variables, $\eta_{(\mathrm{t,r})}$, satisfy $\langle \eta_{(\mathrm{t,r})}(t)\rangle=0$ and $\langle\eta_{(\mathrm{t,r})}(t) \eta_{(\mathrm{t,r})}(t')\rangle=2D_{(\mathrm{t,r})}\delta(t-t')$, where $\langle . \rangle$ denotes ensemble average. We use $D_\mathrm{t}=1$ and $D_\mathrm{r}=3$ in numerical simulations. The particles interact with a pairwise additive force, $\boldsymbol{F}_i[\textbf{r}^N] = \sum_{j\ne i}^N F(r_{ij})\boldsymbol{\hat{r}_{ij}}$, where $r_{ij}=|\textbf{r}_i-\textbf{r}_j|$, $F(r)$ is assumed to be short-ranged and repulsive, and $\boldsymbol{\hat{r}}$ denotes unit vector.In all the simulations, the system of ABPs interact through a WCA potential\cite{Weeks1971} given by 
\begin{equation}
U(r)=\begin{cases} 4\epsilon \left [\left (\frac{\sigma}{r} \right )^{12}+\left(\frac{\sigma}{r} \right )^{6} \right]+\epsilon \quad r < 2^{1/6}\sigma \\ 
0 \quad r \ge 2^{1/6}\sigma
\end{cases}
\end{equation}
where we set the energy scale, $\epsilon$, and lengthscale, $\sigma$, to be 1. From this potential the force is given by $F(r)=-\nabla U(r)$.
All simulation results are presented in Lennard Jones units with time in units of $\tau_\mathrm{LJ}=\sigma^2/( D_t)$ and currents, $J$,  in units of $\sigma/\tau_\mathrm{LJ}$.  Our simulations are in two-dimensions with a domain of 100$\sigma$ x $100\sigma$ and periodic boundary conditions. We used particle numbers of $N=1000$, $ 3000$, and $5000$, which corresponds to densities of $\rho=0.1$, $0.3$, and $0.5$. We used a second order Stochastic Runge Kutta algorithm\cite{Branka1999} with a timestep of $\delta t=10^{-5}\tau_\mathrm{LJ}$. All data presented were computed with 2-3 independent simulations, with a total simulation time between 500-5000 $\tau_\mathrm{LJ}$, including 200 $\tau_\mathrm{LJ}$ of equilibration.

The time integrated current for particle $i$ is defined as
\begin{equation}
\label{Eq:J}
\pmb{J}_i=\frac{1}{t}\int_{0}^{t} d{t'}\,  \dot{\textbf{r}}_i(t') = \frac{\textbf{r}_i(t) -\textbf{r}_i(0)}{t} , 
\end{equation}
where the observation time, $t$, is assumed to be large. The total current for all $N$ particles in the system is $\boldsymbol{J}^N =\{\pmb{J}_{1},\dots,\pmb{J}_{N} \}$. To characterize the statistics of $\boldsymbol{J}^N$ in the long time limit, we aim to compute its LDFs. We define a generating function, 
\begin{align}
\hat{P}(\boldsymbol{\lambda},\boldsymbol{r}^{N},\boldsymbol{\theta}^N,t)=\int d\boldsymbol{J}^{N}\, P(\boldsymbol{r}^N,\boldsymbol{\theta}^N,\boldsymbol{J}^{N},t)e^{ t\boldsymbol{\lambda} \cdot \boldsymbol{J}^{N}}
\end{align}
with ${P(\boldsymbol{r}^N,\boldsymbol{\theta}^N,\boldsymbol{J}^N,t)}$ being the joint distribution of observing all of the particles in a particular position, orientation, and total integrated current, at time $t$. The vector $\boldsymbol{\lambda}$ is conjugate to the current vector and exponentially reweights ${P(\boldsymbol{r}^N,\boldsymbol{\theta}^N,\boldsymbol{J}^N,t)}$.

The time evolution of the generating function is given by
\begin{equation}
\label{Eq:LSO}
\frac{\partial \hat{P}(\boldsymbol{\lambda},\boldsymbol{r}^{N},\boldsymbol{\theta}^N,t)}{\partial t}=L^N_{\lambda}\hat{P}(\boldsymbol{\lambda},\boldsymbol{r}^{N},\boldsymbol{\theta}^N,t)
\end{equation}
that defines the Lebowitz-Spohn operator~\cite{Chetrite2013,Lebowitz1998,Garrahan2010}. This operator has two pieces, $L^{N}_\lambda=L^{N}_{0}+\Delta L^{N}_\lambda$, where
 \begin{equation}
 \label{Eq:Ltilt1}
L^{N}_0=\sum_{i=1}^N \left ( \boldsymbol{F}_i\left [\textbf{r}^N(t)\right ]+{\vo}\boldsymbol{e} \left [ {\theta_i}\left(t\right)\right ]+D_\mathrm{t}{\boldsymbol{\nabla}}_i   \right ) \cdot{\boldsymbol{\nabla}_i}+D_\mathrm{r}\partial_{\theta_i}^{2}\, ,
\end{equation}
is conservative and whose adjoint gives the Fokker Planck operator, and the piece dependent on $\boldsymbol{\lambda}$
\begin{equation}
 \label{Eq:Ltilt2}
\Delta L^{N}_\lambda=\sum_{i=1}^N \left ( \boldsymbol{F}_i\left [\textbf{r}^N(t)\right ]+\vo \boldsymbol{e} \left [ {\theta_i}\left(t\right)\right ]+ 2D_\mathrm{t} {\boldsymbol{\nabla}_i} + D_\mathrm{t}\boldsymbol{\lambda} \right )\cdot \boldsymbol{\lambda}
\end{equation}
does not conserve probability. The spectrum of $L^{N}_\lambda$ is generally complex, but its largest eigenvalue is guaranteed to be real, and whose dependence on $\boldsymbol{\lambda}$ yields the cumulant generating function (CGF) for the current. 

Within this framework we can naturally describe two limiting cases. First, we can consider the statistics of the total system current defined as the sum over the individual particle currents, by setting $\boldsymbol{\lambda}=\lambda \cdot \mathds{1} $ where $\lambda$ is a scalar parameter and $\mathds{1}$ the identity.  However, this case is trivial because the sum of the interparticle force in Eq.~\ref{Eq:Ltilt2} vanishes, decoupling the equation into a sum of $N$ independent equations. In this case, the total current CGF is equivalent to $N$ times the CGF for a single ABP.  Alternatively, we can consider the current statistics of a single tagged ABP, subject to the interactions of the surrounding particles. This is done by setting $\boldsymbol{\lambda}$ to be a vector with a single nonzero element, $\boldsymbol{\lambda} =\{0,0,0,\cdot \cdot \cdot, \lambda,\cdot \cdot \cdot ,0,0 ,0\}$. This second case contains the first in the limit of low density, and provides additional information on the dependence of current fluctuations on interactions. In the following we will consider the second definition.

\begin{figure*}[t]
\begin{center}
\includegraphics[width=16cm]{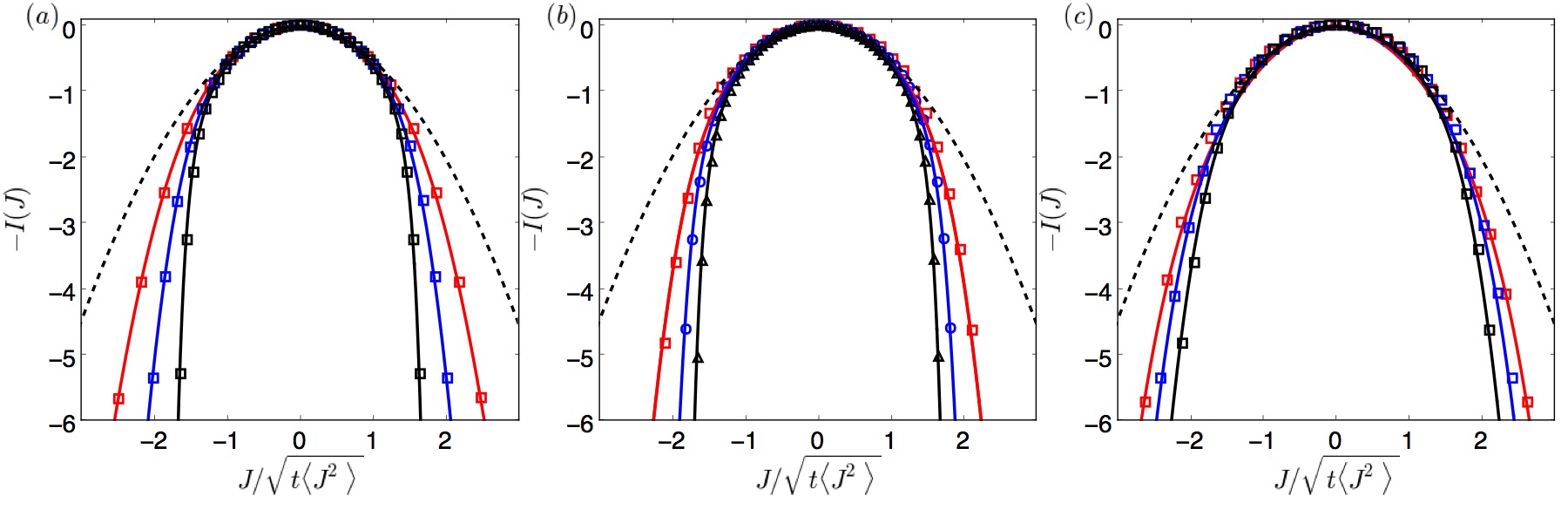}
\caption{Comparison between the analytical rate function and its numerical evaluation. (a) Rate functions for $\rho=0$ and $\vo=$ 5 (red), 10 (blue), and 60 (black). (b) Rate functions for $\rho=0.1$ and $\vo=$ 10 (red), 30 (blue), and 60 (black). (c) Rate functions for $\vo=10$ with $\rho= 0.1$ (red), 0.3 (blue), and 0.5 (black). Shown are the Legendre transforms of Eq.~ \ref{Eq:Psi} (solid lines), numerical simulations (symbols) and reference Gaussian (dashed line). }
\label{Fi:1}
\end{center} 
\end{figure*}

In order to calculate the CGF for a tagged particle, we first introduce a weighted many body expansion that follows from a BBGKY-like hierarchy~\cite{Hansen1977}. Specifically, we define an $n$-particle reduced generating function 
\begin{align}
\label{Eq:Pofn}
\hat{P}^{(n)}(\lambda, \boldsymbol{r}^n,\boldsymbol{\theta}^n,t)=&\\
\frac{N!}{(N-n)!} & {\int}{\int}d{\boldsymbol{r}^{(N-n)}}d{\boldsymbol{\theta}^{(N-n)}}\hat{P}(\lambda,\boldsymbol{r}^N,\boldsymbol{\theta}^N,t) \, ,\nonumber
\end{align}
which, when introduced into Eqs.~\ref{Eq:LSO}-\ref{Eq:Ltilt2}, will result in a set of coupled evolution equations for different $\hat{P}^{(n)}$'s. 
We can close the single particle equation with the two-particle generating function, decomposed as
\begin{equation}
\label{Eq:Gofr}
g_\lambda(\boldsymbol{r},\theta,\boldsymbol{r'},\theta',t) = \frac{\hat{P}^{(2)}(\lambda, \boldsymbol{r},\theta,\boldsymbol{r'},\theta',t)}{\hat{P}^{(1)}(\lambda,\boldsymbol{r},\theta,t)\hat{P}^{(1)}(\lambda,\boldsymbol{r'},\theta',t)}
\end{equation}
where $g_\lambda(\boldsymbol{r},\theta,\boldsymbol{r'},\theta',t)$ is the pair distribution function conditioned on a given current through $\lambda$~ ~\cite{Appert-Rolland2008,Hedges2009}. This function can be simplified when the  system is in a homogeneous steady-state and assuming that it does not depend on the difference in orientations between particles. In this limit,  $g_\lambda(\boldsymbol{r},\theta,\boldsymbol{r'},\theta',t) \approx g_\lambda(r,\phi)$, where $\phi$ is the angle of the displacement vector of two particles relative to the orientation of the particle at the origin. This closure to the many-body hierarchy was introduced previously for the case of $\lambda=0\,$~\cite{Speck2015,Bialke2013,Hancock2017,Wittkowski2017}.

The equation of motion for the single particle generating function will depend on the average interparticle force. We can decompose this force into components in the parallel and perpendicular direction of the self-propulsion, however this will result in an average force that depends on both the relative angle between the interparticle displacement vector and the tagged particle's orientation. Following Speck et al.\cite{Speck2015} if we approximate the component perpendicular to the orientation as that parallel to the surface of the particle, this will uncouple these two terms allowing for the expansion to be closed. This approximation to the perpendicular force is exact in the limit of passive particles, where there are no orientational correlations, and is numerically accurate when the parallel component is much larger
than the perpendicular component, as occurs for $\vo > 1$. 

As we consider only homogeneous systems, for notational simplicity and without loss of generality we restrict our attention to currents in just the $x$ direction.  Under these assumptions, we obtain the evolution operator for the single particle generating function for currents, 
  \begin{equation}
  \label{Eq:Ltilts}
L_{\lambda}= V_\lambda(\rho)\cos(\theta)(\partial_x +\lambda)+ D_\mathrm{t}(\rho)(\partial_x+\lambda)^2+D_\mathrm{r}\partial_{\theta}^{2},
\end{equation}
that has the same drift-diffusion form as an independent ABP, but with renormalized effective propulsion speed, $V_\lambda(\rho)$, and translational diffusion constant, $D_\mathrm{t}(\rho)$, where $\rho$ is the local density, which in the homogeneous assumption is taken as the bulk density. The adjoint of the operator in Eq.~\ref{Eq:Ltilts} evaluated at $\lambda=0$ yields the propagator for the single particle density.

Both $V_\lambda(\rho)$ and $D_\mathrm{t}(\rho)$ in principle depend on
$\rho$, $\vo$ and $\lambda$, through the state-dependent pair correlation function.
Within this force decomposition, 
$D_\mathrm{t}(\rho)$ is the diffusion coefficient for a system of interacting passive Brownian particles and we have found that for the conditions we study, $D_\mathrm{t}(\rho)$  can be approximated by the mean field form, $D_\mathrm{t}\left(\rho\right)\approx D_\mathrm{t}\left(1-\rho\right)$~\cite{Hancock2017}. 
The effective propulsion speed takes the form $V_\lambda(\rho)=\vo-\rho\zeta_\lambda(\rho)$ where $\rho\zeta_\lambda(\rho)$ is an effective drag. This drag is given by an integral over the interparticle force
\begin{equation}
\label{Eq:zeta}
\zeta_\lambda \left(\rho\right)=\int_{0}^{\infty} d{r}\, \int_{0}^{2\pi}d{\phi}\,r \cos(\phi) g_\lambda(r,\phi)F(r) 
\end{equation}
weighted by the pair distribution function. This coefficient describes the decrease in the effective velocity of a tagged particle due to the increased density of impenetrable particles in the direction of self-propulsion\cite{Speck2015}. In the following, we take $\rho\zeta_\lambda(\rho)$ as input for our evaluation of the CGF,  though simple approximations to $\zeta_\lambda(\rho)$ exist in specific limits~\cite{SI}.

Using these definitions, we are able to solve for the CGF for this effective single particle description of the system by the largest eigenvalue of the equation $L_{\lambda}\nu_\lambda=\psi(\lambda)\nu_\lambda$, with $\psi(\lambda)$ being the CGF and $\nu_\lambda$ its corresponding right eigenvector. The solution is given by the zeroth characteristic function of Mathieu's equation~\cite{Abramowitz1965}, with a representation for small $\lambda$ given by the expansion,
\begin{align}
\label{Eq:Psi}
\psi\left(\lambda\right)&=D_\mathrm{t}\left(\rho\right)\lambda^2\\
&+D_\mathrm{r} \left [\frac{z_{\lambda}^2(\rho)}{2}-\frac{7 z_{\lambda}^4(\rho)}{32}+\frac{29 z_{\lambda}^6(\rho)}{144}\right ] +\mathcal{O}(\lambda^8) \nonumber
\end{align}
with $z_{\lambda}(\rho)=V_\lambda(\rho) \lambda/ D_\mathrm{r}$. The CGF is symmetric about $\lambda=0$ as a consequence of spatial inversion symmetry and retains all even powers, alternating in sign. The terms up to second order in $\lambda$ represent the Gaussian contribution. All higher order terms in $\lambda$ give the Non-Gaussian behavior. Specifically, the excess Kurtosis, which is a common metric for Gaussian deviations is given by the term that is quartic in $\lambda$. For passive particles $z_{\lambda}(\rho)=0$, and $\psi\left(\lambda\right)$ reduces to that for Brownian motion with an effective diffusion constant, $D_\mathrm{t}(\rho)$. In the limit that the particles are non-interacting, or $\rho\rightarrow 0$, our results reduce to those obtained previously ~\cite{Pietzonka2016}.

Given the CGF, the rate function for current fluctuations can be computed from the Legendre-Fenchel transform, $I\left(J\right) = \max_{\lambda}[\lambda J-\psi\left(\lambda\right)]$, where $I(J)$ is minus the logarithm of the probability of $J$ divided by the observation time.
Figure \ref{Fi:1} shows the rate functions computed from the cloning algorithm \cite{Lecomte2007,Giardina2006} and predictions from $\psi(\lambda)$. For a variety of different $\rho$'s and $\vo$'s we find quantitative agreement between the analytical result and the simulations. 
Small fluctuations around $J=0$ are Gaussian as expected, but larger fluctuations are markedly non-Gaussian, revealing fluctuations that are more rare than anticipated from the time intensive variance, $t\langle J^2\rangle$. The deviations from Gaussian behavior become more distinct with increasing $\vo$ and decreasing $\rho$ as highlighted in Fig.~\ref{Fi:1} by scaling the current by $\sqrt{t\langle J^2\rangle}$. 

We can gain insight into the shape of $I(J)$ by constructing an auxiliary process that generates the same ensemble of trajectories in the long time limit as the original model conditioned on a given current~\cite{Chetrite2015}. The auxiliary process is a transformation of the Lebowitz-Spohn operator,
\begin{align}
\mathcal{L}_{\lambda}=&\,  \nu_\lambda\left(\theta\right)^{-1}L_\lambda \nu_\lambda\left(\theta\right) -\psi\left(\lambda\right)\nonumber \\
=&\, L_0+2D_\mathrm{t}(\rho)\lambda\partial_{x} +2 D_\mathrm{r} \partial_\theta \ln \nu_\lambda\left(\theta\right) \partial _{\theta} 
\end{align}
which leaves the diffusion terms unmodified, but adds additional drift terms in that restore normalization. The auxiliary process for a tagged particle is
 \begin{equation}
 \label{Eq:Aux1}
\dot{\textbf{r}}(t) =\boldsymbol{F}\left [\textbf{r}^N(t)\right ]+{\vo}\boldsymbol{e}\left [ {\theta}\left(t\right)\right ]+2D_\mathrm{t}(\rho)\lambda \hat{\boldsymbol{x}}+\boldsymbol{\eta}_{\mathrm{t}}\left(t\right)
\end{equation}
where the added force is a constant proportional to $\lambda$ in the $x$ direction. The equation of motion for the orientation includes a force, $F_{\lambda}(\theta)=2 D_\mathrm{r} \partial_\theta \ln \nu_\lambda\left(\theta\right)$, which for small $\lambda$ is
\begin{equation}
 \label{Eq:Aux2}
  \dot{\theta}_i \left(t\right) = -2 V_\lambda(\rho) \lambda \sin(\theta)+{\eta_{\mathrm{r}}\left(t\right)}
\end{equation}
where the force has an amplitude that depends on $\lambda$ directly and through the $\lambda$-dependent drag coefficient, $\zeta_\lambda(\rho)$. The exact form of $\nu_\lambda\left(\theta\right)$ can be evaluated by basis set expansion of Eq. \ref{Eq:Ltilts} \cite{Pietzonka2016}, and $\zeta_\lambda(\rho)$, can be evaluated self-consistently using the generalized variational principle of Ref.~\cite{ray2017exact}, as direct evaluation of $g_\lambda(r,\phi)$ is exponentially difficult. From the result in Eq. \ref{Eq:Aux1}, we could easily generalize to two dimensional bias by adding one additional term of $2D_\mathrm{t}(\rho)\lambda \hat{\boldsymbol{y}}$. 

\begin{figure}
\begin{center}
\includegraphics[width=8.5cm]{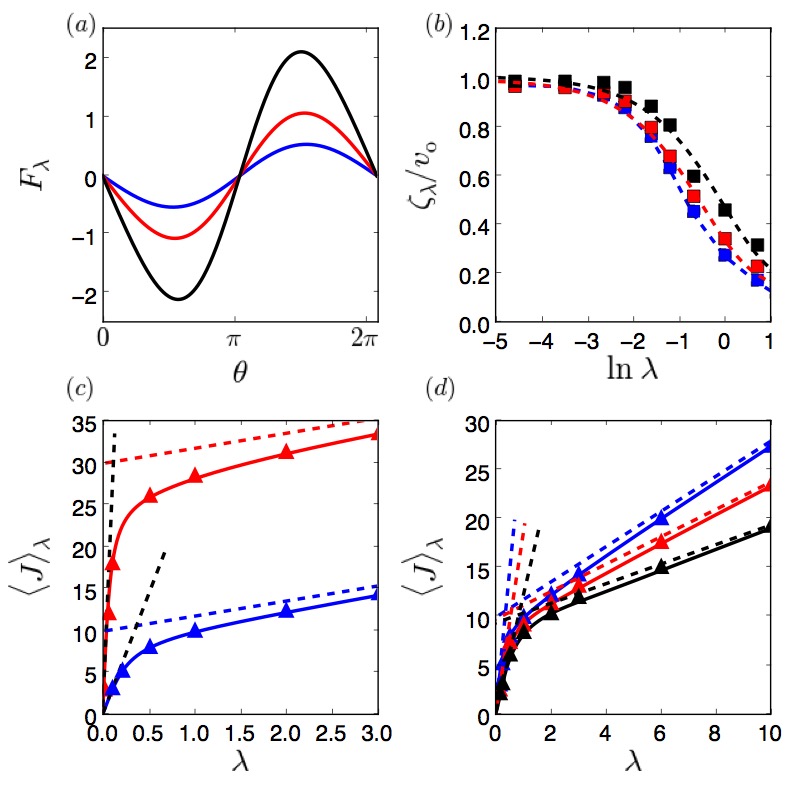}
\caption{
Analysis of the auxiliary process. (a) The effective forces for $\vo$=10, $\rho=0.1$ and $\lambda=0.1$ (blue), 0.3 (red), and 0.5 (black). (b) Damping coefficient, $\zeta_{\lambda}$, as a function of $\lambda$ for $\vo$=10 and $\rho= 0.1$ (blue), 0.3 (red), and 0.5 (black). Dashed lines are a guide to the eye. (c,d) The average current from the auxiliary process. (c) $\langle J \rangle_\lambda$ for $\rho=0.1$ and $\vo$=10 (blue) and 30 (red). (d) $\langle J \rangle_\lambda$ for $\vo$=10, and $\rho= 0.1$ (blue), 0.3 (red) and 0.5 (black). The symbols are the results from simulations. The solid lines represent the derivative of the CGF and the dotted lines represent its limiting behavior.}
\label{Fi:Fig2}
\end{center} 
\end{figure}

Within the auxiliary process, rare large currents result from the effective force that confines the orientation of the active particle to a given direction.  This is shown in Fig.~\ref{Fi:Fig2}a), where the force has stable points at $\theta=0$ and $\pi$, depending on the sign of $\lambda$, with an amplitude that grows with increasing $\lambda$~\cite{Pietzonka2016}.  Additionally, the effective drag from the surrounding particles is reduced with increasing magnitude of $\lambda$. Shown in Fig. ~\ref{Fi:Fig2}b) is $\zeta_\lambda(\rho)$ computed from the molecular simulations for a variety of densities. From inversion symmetry, $\zeta_\lambda(\rho)$ is an even function about $\lambda=0$, and we find for small values of $|\lambda |$, it decreases quadratically.  In the large $|\lambda |$ limit, we find that $\zeta_\lambda(\rho)$ decreases exponentially to 0, resulting in effectively free particle evolution. This decrease reflects the reduced probability of particle collisions in the direction of the bias and the onset of hyperuniformity~\cite{Torquato2003,Jack2015}.

Derivatives of $\psi(\lambda)$ provide the cumulants of $J$,  $d^n \psi(\lambda)/d \lambda^n = C^n_\lambda$, with the first, $C^1_\lambda= \langle J \rangle_\lambda$, yielding the average current, and the second, $C^2_\lambda= t \langle (J-\langle J\rangle_\lambda)^2 \rangle_\lambda$, its variance. When evaluated at $\lambda=0$, these are cumulants of the original model, but for $\lambda \ne 0$, these report on rare fluctuations into the tails of $I(J)$. Shown in Figs.~\ref{Fi:Fig2}c) and d), are the average currents computed at finite $\lambda$, from the exact solution of the eigenvalue equation and from evaluating Eq.~\ref{Eq:J} directly from simulations of the auxiliary process defined in Eqs.~\ref{Eq:Aux1} and \ref{Eq:Aux2}. For small $\lambda$, the current increases linearly from 0 with a slope set by variance at $\lambda =0$, as is expected from linear response. For large $\lambda$, the system exhibits nonlinear response, manifesting the non-Gaussian current fluctuations. In this limit, the slope decreases dramatically. The asymptotic limit of this secondary response is given by an offset of $\vo$ and slope that depends only on the $D_\mathrm{t}(\rho)$. The origin of this dependence is clear from the auxiliary process. For large $|\lambda|$, $V_{\lambda}(\rho)\rightarrow \vo$, and the force on the orientation suppresses angular fluctuations. Analysis of this limit provides an asymptotic form for $\psi(\lambda)$,
\begin{align}  
\psi\left(\lambda\right)= D_\mathrm{t}\left(\rho\right)\lambda^2 \pm \vo \lambda \,,  \quad\,  \lambda \rightarrow \pm \infty\, ,
\end{align}
which shows that the tails of the $I(J)$ are given by an effective Gaussian with mean, $\vo$, a much smaller variance than at $\lambda=0$.  We note a simple effective temperature mapping between Brownian particles and ABPs would not explain the observed non-Gaussian fluctuations or concomitant secondary response~\cite{Steffenoni2017,Solon2015,Takatori2015,Palacci2010}. Further, this qualitative behavior of two different effective diffusion constants for small and large fluctuations has been observed recently in active biopolymers\cite{Wang2012,Stuhrmann2012}.

\begin{figure}[t]
\begin{center}
\includegraphics[width=8.5cm]{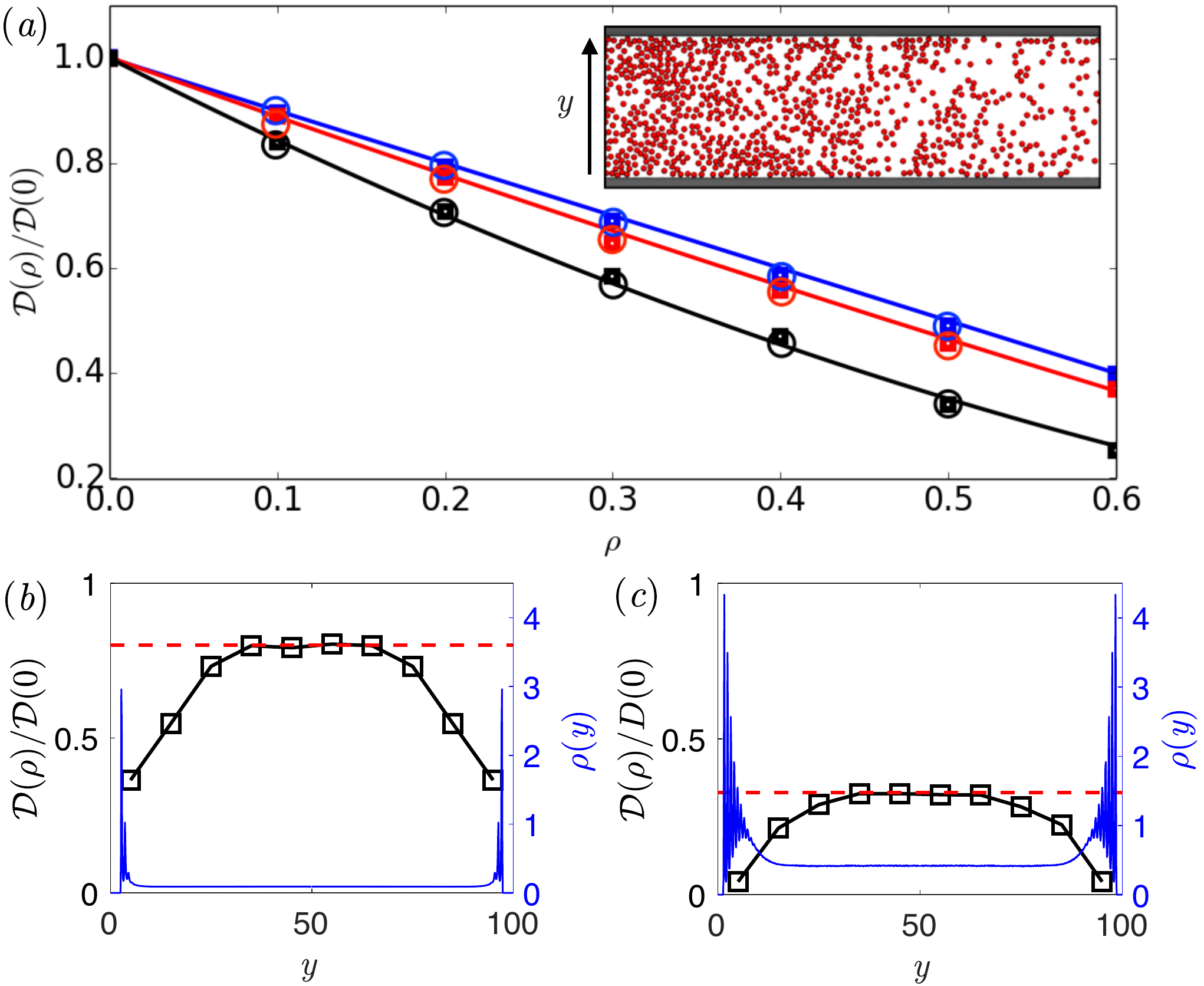}
\caption{(a)Density dependent effective diffusion constant for $\vo= 0$ (blue), 1 (red), and 60 (black). The symbols are the results from simulations. The squares are from the mean squared displacement, open circles are from an imposed density gradient and the solid lines from Eq.~\ref{Eq:Diff}. Inset shows a snapshot of the simulation with an imposed density gradient and bulk density of $\rho=0.2$. The density and diffusion profiles along the width of the channel for (b) $\vo= 60$ $\rho=0.1$, and (c)  $\vo=60$ $\rho=0.5$. The dashed red lines are the predicted values.}
\label{Fi:Fig3}
\end{center} 
\end{figure}

We conclude with a discussion of the second cumulant,
\begin{equation}
\label{Eq:Diff}
C^2_0(\rho) =2D_\mathrm{t}(\rho)+\frac{V_{0}^2(\rho)}{D_\mathrm{r}}\equiv 2\mathcal{D}(\rho)\, ,
\end{equation}
that we define as twice a collective diffusion constant, $\mathcal{D}(\rho)$. In Fig.~\ref{Fi:Fig3}, numerical results obtained from simulations of the mean-squared displacement divided by a diffusive observation time are plotted in excellent agreement with predictions from Eq.~\ref{Eq:Diff}. This form of the diffusion coefficient has been shown to agree with simulations previously, and was derived by a moment expansion of the joint position and orientation distribution~\cite{Stenhammar2014,Chakraborti2016,Fily2012,Cates2015}. This density dependence of $\mathcal{D}(\rho)$ was shown by others~\cite{Bialke2013,Speck2015,Solon2015a} to correctly predict the spinodal instability signaling the onset of motility induced phase separation~\cite{SI}. 

The current fluctuations encoded by $\mathcal{D}(\rho)$ provide the response of a hydrodynamic current, $J_\rho$, generated from a slowly varying spatial density, $\rho(x)$. From the Kramer's Moyal expansion~\cite{Risken1996}, $J_\rho$ can be generally expressed as a gradient expansion
\begin{align}
 J_\rho = -\sum_{n=1}^\infty \frac{(-1)^n}{n !} \partial_x^{n-1} M^n[\rho(x)] \rho(x)
 \end{align}
 where $M^n[\rho(x)]$ is the local density-dependent $n$th moment of the current, $\langle (J-\langle J \rangle)^n \rangle$.  To first order, the mass current is linear in the density gradient and is given by Fick's law, $J_\rho \approx -\mathcal{D}(\rho)(\partial \rho/\partial x)$, where $\mathcal{D}(\rho)$ is the proportionality constant relating the current to the gradient resulting from identifying the second moment with the second cumulant. Since for small average currents we have  $\langle J \rangle_\lambda \approx 2 \mathcal{D}(\rho) \lambda$, which shows that at linear response, $\lambda$ can be related to an affinity for this nonequilibrium system. We have computed $\mathcal{D}(\rho)$  from $-J_\rho /(\partial \rho /\partial x)$ by simulating an open channel in contact with two reservoirs.  As shown in Fig. 3a), we find good agreement with $\mathcal{D}(\rho)$ computed in this way and from $\psi(\lambda)$.
 
 The presence of the walls result in a nonuniform density distribution in the direction orthogonal from the walls, with decay lengths that increase with $\rho$ and $v_\mathrm{o}$ as found previously
 \cite{Li2009,Elgeti2015,Elgeti2013,Fily2014a,Yang2014,Elgeti2015a,wagner2017steady}. 
 We simulate channels that are wide enough to be bulk-like in the center of the channel. A boundary layer that results from the particle accumulation at the wall has a reduced local diffusion constant parallel to the walls. This is due to the increased density and the correlation of the self-propulsion vector and the normal force of the wall. Figures 3b) and c) show two representative diffusivity and density profiles at large  $\vo$ and small and large $\rho$. The range over which the diffusion constant reaches its bulk plateau value determines the effective width of the channel for computing the response due to the density gradient. Additional simulation details and discussion of the density layering near the walls can be found in the SI \cite{SI}.

 From $\psi(\lambda)$, we have access to all moments of $J$, and together with its $\rho$ dependence this framework allows us to quantify higher order responses~\cite{SI} that are not naturally considered in standard field theoretic treatments of ABPs\cite{Speck2015,Bialke2013,Hancock2017,Stenhammar2014,Solon2015}.
While our focus has been on ABPs, the framework we have presented is general and allows for the quantification of current fluctuations, and the calculation of transport coefficients for continuous interacting systems. 
For ABPs, we found that large current fluctuations near the mean are not representative of rarer fluctuations which are restricted as a result of coherent active movement. 
These specific results are consistent with deviations from Gaussian behavior that have been reported in recent experimental studies of active colloids\cite{Zheng2013,Io2017}. 
Furthermore, our results may explain the center of mass motion of cellular biopolymers that exhibit two types of transport characterized by two different diffusion constants for small and large fluctuations \cite{Wang2012,Stuhrmann2012}. 

While other types of active matter such as active rods, biopolymers and bacteria behave like ABPs in specific limits, each also adds complexities not accounted for in our present treatment. These include  additional interactions like those due to preferential alignment or hydrodynamics, and internal degrees of freedom[41, 81-87]. Developing many-body closures appropriate for these contexts can take inspiration from previous work on molecular fluids like the reference interaction site model \cite{Chandler1972} and polymer reference interaction site model equations \cite{Curro1987}, and is an interesting direction for further study.
 Finally, while we have focused on current fluctuations, our development of the weighted many body expansion provides a way to calculate the LDFs of other relevant quantities for nonequilibriums systems such as activity \cite{Whitelam2018,Jack2015,Pitard2011,Hedges2009,Garrahan2007,Katira2017},  entropy production \cite{Mandal2017,Fodor2016,Mehl2008,Speck2012} and density \cite{Derrida2001,Chakraborti2016} that are currently difficult to estimate. 

\textbf{Acknowledgements:} D.T.L was supported by UC Berkeley College of Chemistry. T.G.P was supported by the Kavli Energy NanoSciences Institute. We thank Katherine Klymko for help in setting up simulations.

\pagebreak
\widetext
\begin{center}
\textbf{\large Current fluctuations of interacting active Brownian particles: \\Supplementary Information}
\end{center}
\setcounter{equation}{0}
\setcounter{figure}{0}
\setcounter{table}{0}

\author{Trevor Grand Pre}
\affiliation{%
Department of Physics, University of California, Berkeley 
}

\author{David T. Limmer} \email{dlimmer@berkeley.edu}
\affiliation{%
Department of Chemistry, University of California, Berkeley 
}
\affiliation{%
Kavli Energy NanoSciences Institute, University of California, Berkeley 
}\affiliation{%
Lawerence Berkeley National Laboratory, University of California, Berkeley 
}

\section{Simulations details}
The current was estimated by  
\begin{align}
\boldsymbol{J}_i=&\, \frac{\mathbf{r}_{i}(\Delta t)-\mathbf{r}_{i}(0)}{\Delta t}. 
\end{align}
which is the total displacement of particle $i$ over the observation time, $\Delta t$. The observation time is chosen to be in the long time limit where all current distributions for that time and larger have a large deviation form.  We find $\Delta t=10 \tau_\mathrm{LJ}$ sufficient for all conditions studied. The numerical distributions in Fig. 1 of the main text were obtained by taking the bin edges of the histogram of currents. The analytical results were calculated by taking the numerical Legendre transform of the CGF, $\psi\left(\lambda\right)$. 
We calculate the CGF using a Fourier-Bloch decomposition of the Lebowitz-Spohn operator. We find that using a basis of 25 cosine functions is sufficient to converge $\psi(\lambda)$ for all conditions studied~\cite{Pietzonka2016}. 

In Fig.  2a), we show the calculation of the effective force given by $F_{\lambda}=2 D_r \partial_\theta \ln \nu_{\lambda}\left(\theta\right)$, where $\nu_{\lambda}\left(\theta\right)$ is the right eigenvector of the CGF.
This effective force is fit to a sum of sines to get a functional form and then used in the equations of motion to generate an auxiliary dynamics that realize the large currents ~\cite{Chetrite2015}. We integrate these auxiliary equations of motion and measure the current according to Eq.  S1. The average currents from simulations are compared with the derivative with respect to $\lambda$ of the CGF, $\psi(\lambda)$.
The diffusion coefficient is estimated by measuring the mean-squared displacement in the diffusive regime:
\begin{equation}
\mathcal{D(\rho)}=\frac{1}{4N}\sum_{i=1}^{N}\frac{| \mathbf{r}_{i}(\Delta t)-\mathbf{r}_{i}(0)|^2}{\Delta t}
\end{equation}
with $\Delta t=10 \tau_\mathrm{LJ}$. 

In order to compute the diffusion coefficient from an imposed density gradient, a long channel connecting two large reservoirs was constructed out of static WCA particles. The dimensions of each reservoir was $100\sigma$ by $150\sigma$ in the $x$ and $y$ directions, and the channel was $200\sigma$ by $100\sigma$, as determined by an insensitivity of the computed diffusion constant to the system geometry. An initial density gradient was generated by initializing the reservoirs with densities $\pm 5 \%$ of the mean target density in the channel, which in all cases was well within the linear response regime. After an initial transient period, a steady-state current was evaluated by counting the number of particles transferred from one reservoir to the other per time and per width of the channel. Statistics were accumulated over a simulation time where the current was linearly proportional to time, typically 3000 $\tau_\mathrm{LJ}$. The gradient was measured away from the entrance of the channel and the diffusion constant was computed by the ratio of the gradient to the flux.

\section{Density layering of ABPs near extended walls}
\begin{figure}[h]
\begin{center}
\includegraphics[width=16cm]{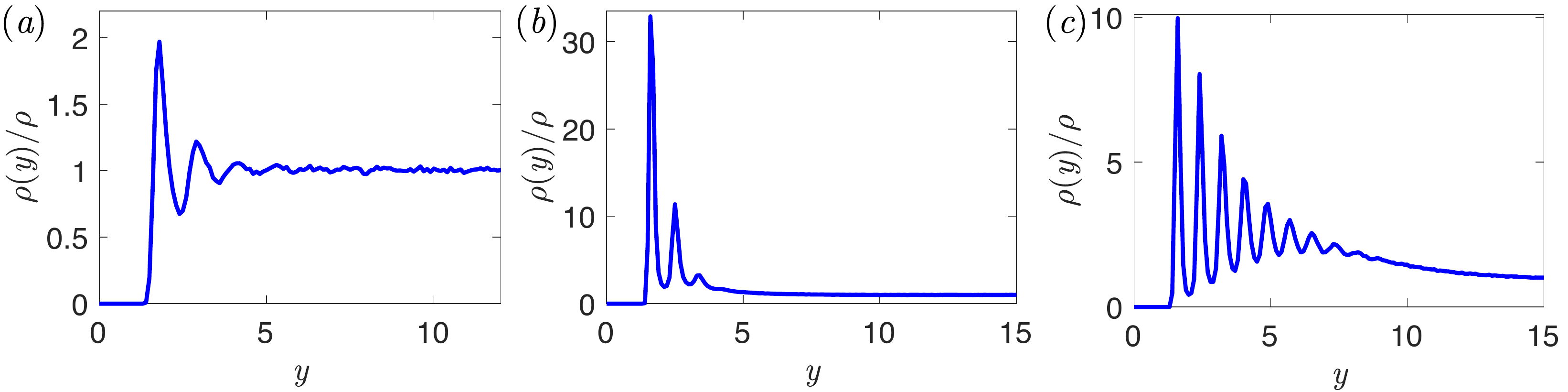}
\caption{The zoomed in density profiles along the width of the channel. (a) The density profile for  $v_0=1$ and $\rho=0.5$. (b) The density profile for $v_0=60$ and $\rho=0.1$. (c) The density profile for $v_0=60$ and $\rho=0.5$}
\label{Fig:SI1}
\end{center} 
\end{figure}
Breaking translational symmetry by placing a wall in the system generically results in density correlations away from the wall. In order to quantify the decay of these correlations to ensure that our transport computations are dominated by the bulk behavior we have evaluated the density profiles away from the wall for a broad range of conditions. These are shown in Fig~\ref{Fig:SI1}.  Figure~\ref{Fig:SI1}a) shows that for $v_\mathrm{o}=1$ and $\rho=0.5$ there is very little accumulation. The profile is like that of a passive low density liquid. Figure~\ref{Fig:SI1}b) shows the density profile for $\vo=60$ and $\rho=0.1$ with a large but narrow first peak showing a small layer of accumulation on the walls due to correlations of the self-propulsion vector with the direction perpendicular to the wall.  Figure~\ref{Fig:SI1}c) shows that for increasing $\vo$ and increasing $\rho$, such as $\vo=60$ and $\rho=0.5$, the layering is enhanced and additionally there is a distinct exponential tail to the density profile. In these most extreme conditions, the correlations still decay on a length scale of 10 $\sigma$, which is much smaller than the width of the channel.

Furthermore, accumulation does not affect the transport measurements because we  set  the  system  up such  that  far  away  from  the  walls,  the  density  is  the  reported  density  for  our  calculation by increasing  the  particle  number of the system  until  the  excess  at  the  walls  is compensated. Additionally, we quantify the boundary layer between the accumulating particles on the walls and the bulk by where  the  local  diffusion  constant  reaches  its  bulk value, and this defines the effective width of the system for computing the transport coefficients from the imposed density gradient.

\section{Current Distribution in the Coexistence region}
\begin{figure}[h]
\begin{center}
\includegraphics[width=8.5cm]{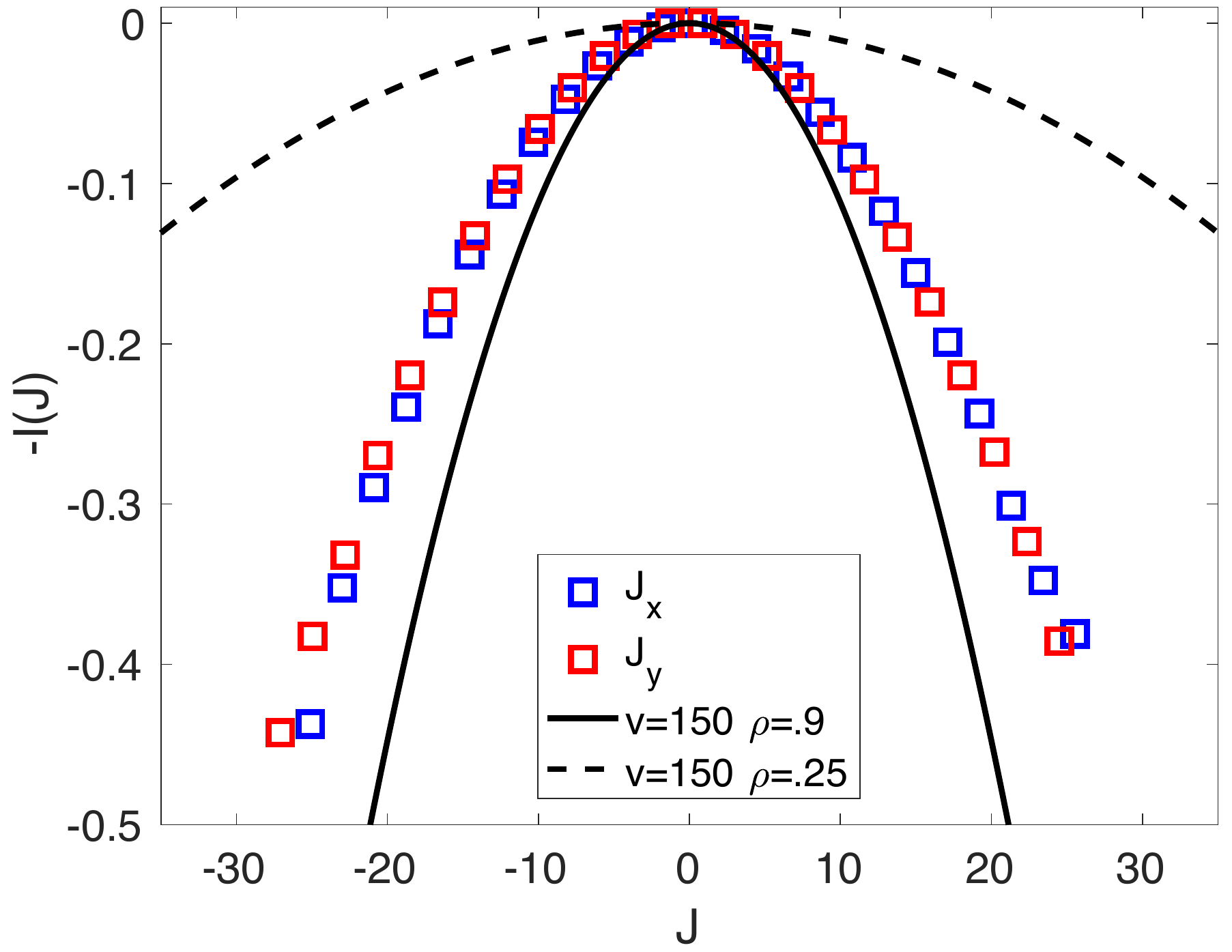}
\caption{The distribution of current fluctuations for $\vo=150$ and $\rho=0.9$ under conditions of phase separation. }
\label{Fig:SI3}
\end{center} 
\end{figure}
The distribution of current under conditions where phase separation occurs is not expected to be predicted by our analytical results since they assume the system is homogeneous. The breakdown of our results for a phase separated system is demonstrated in Fig.~\ref{Fig:SI3}. The current distribution was measured with an observation time of $\Delta t=30\tau_{LJ}$ which is in the long time limit for these system parameters.The system size was $100\sigma$ by  $100\sigma$ in the x and y directions. The total simulation time was $1000\tau_{LJ}$.  
The blue and red squares represent the current fluctuations in the $x$ and $y$ directions. Hence, Fig.~\ref{Fig:SI3} demonstrated that even in a macroscopically heterogeneous system the $x$ and $y$ directions give identical statistics in the long time limit. We also see that for the range simulated the current distribution is unimodal.

To compare this data to our analytical theory, we could use as input the bulk system density to parameterize our CGF. In Fig.~\ref{Fig:SI3}, the black solid line is the prediction for $\vo=150$ and $\rho=0.9$ with $\zeta_0=0.84$. This estimate under-predicts the size of fluctuations. This is because under phase separation conditions, locally the density deviates strongly from its average, resolving into a bimodal distribution\cite{Redner2013}. In the dilute phase, the local density is $\rho=0.25$. If we use that density to parameterize our CGF, we predict the current fluctuations to be much broader than is observed. This prediction is shown in the black dotted line in Fig.~\ref{Fig:SI3}. We cannot use the dense phase density to arrive at a similar prediction, because its local density exceeds 1, which results in a prediction that there are no currents due to crystallization.

\section{Calculation of $\zeta_{\lambda}(\rho)$}
To evaluate $\zeta_{\lambda}(\rho)$ we use molecular simulations and a generalized variational principle~\cite{ray2017exact}. To do this, we first run the auxiliary dynamics with an approximate effective potential derived with $\zeta_0(\rho)$ for $\lambda$ values ranging from 0 to 20. Then, we recalculate $\zeta_{\lambda}(\rho)$ using an approximate $g_\lambda(r,\phi)$ computed from these auxiliary dynamics and Eq.~11. 

We fit the $\zeta_{\lambda}(\rho)$ as a function of $\lambda$, which is then incorporated into the Lebowitz-Spohn operator (Eq. 10 in the main text). This operator is re-diagonalized and its largest eigenvector is used to obtain a better estimate of the effective potential. This process is iterated until $\zeta_{\lambda}(\rho)$ is self-consistent, typically requiring only 2 or 3 iterations. This condition does not ensure an optimal estimate of the CGF, as the Lebowitz-Spohn operator is not Hermitian, but it does enforce stationarity. 


The drag coefficient has a number of known limiting forms. By construction, for independent particles or $\rho \rightarrow 0$, the drag $\rho \zeta_{\lambda}(\rho)\rightarrow 0$. We find that for all $\vo$, this approach is linear in $\rho$. Similarly, for $\vo \rightarrow 0$, the system becomes isotropic and $\zeta_{\lambda}(\rho)\rightarrow 0$. For large $\vo$, $\zeta_{0}(\rho) \approx \vo/\rho^*$ where $\rho^* \approx 1.2$, corresponding to a effective closed-packing density. We find that for $\vo > 30$, this approximation is within 1\%, but even for $\vo \approx 5$, this form is within $10\%$ of the computed value.

\section{Comparison of pair distribution function at finite $\lambda$}
\begin{figure}[h]
\begin{center}
\includegraphics[width=8.5cm]{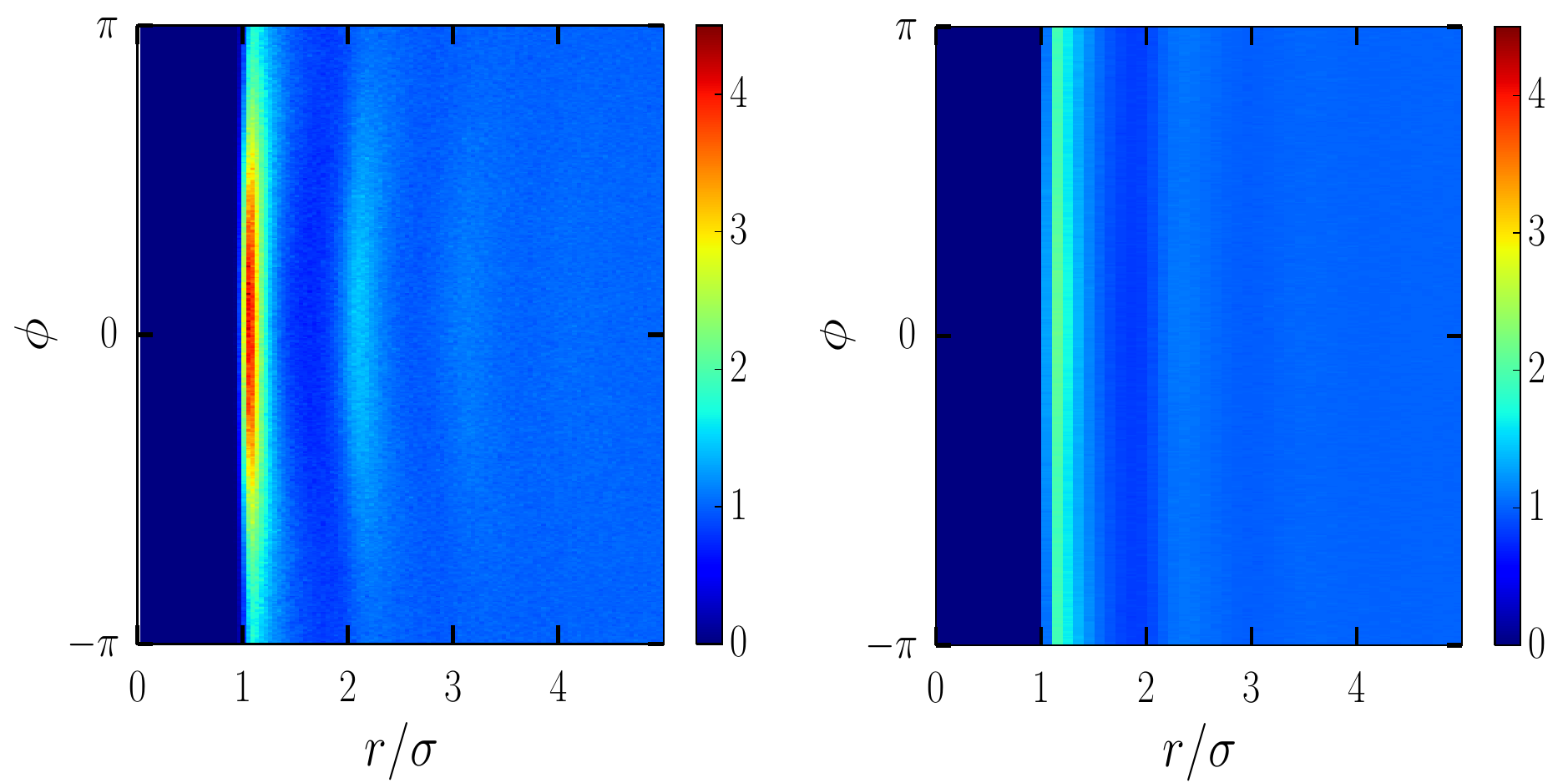}
\caption{Dependence of the pair distribution function on $\lambda$. Pair distribution function for $\lambda=0$ (left), and $\lambda=20$ (right) for $\vo=10$ and $\rho=0.5$.}
\label{Fig:SI}
\end{center} 
\end{figure}

Shown in Fig.~\ref{Fig:SI} are the pair distribution functions, $g_\lambda(r,\phi)$, computed at $\lambda=0$ and $\lambda=20$ for $\vo=10$ and $\rho=0.5$. At $\lambda=0$, $g_0(r,\phi)$ is anisotropic, meaning that there are more particles in front of a tagged particle and less behind. This imbalance and the bow wave-like structure of the correlations result in a nonzero value of $\zeta_\lambda(\rho)$. This is the same coefficient that has been calculated previously and describes the linear decrease of the hydrodynamic velocity with density due to interactions ~\cite{Cates2015,Speck2015,Stenhammar2014}. At large $\lambda$, $g_\lambda(r,\phi)$ becomes uniform in $\phi$ and accordingly $\zeta_\lambda(\rho)$ is 0. In addition, there is a reduction in the intensity of the peaks, most significantly in the first peak that drops from a value of 4.2 to 1.5. These correlations at large $\lambda$ are like those of a low density system of passive Brownian particles, which have no angular dependence, despite the significant net particle drift in one direction.

\section{Relation between $\lambda$ and affinity}

If we include the density dependence of the effective diffusion constant, then we arrive at an expression for the hydrodynamic current,
 \begin{align}
 J_{\rho}\approx -\left [\frac{ \rho V_0(\rho)}{D_r}\frac{\partial V_0(\rho)}{\partial \rho}+\mathcal{D}\left(\rho\right) \right ] \frac{\partial \rho}{\partial x}
  \end{align}
 which includes an extra term omitted in the expression in the main text for Fick's law. We have neglected the density dependence of $D_t(\rho)$, because it is small compared $V_0(\rho)$. The first term on the right is negative, so this expression provides a prediction for phase separation used previously~\cite{Cates2015,Speck2015,Fily2012,Fily2014,Stenhammar2014}.
 
 We can define a density dependent effective chemical potential, $\mu_0(\rho)$, by rewriting the hydrodynamic current as
\begin{equation}
J_{\rho}=-\mathcal{D}(\rho)\rho \frac{d \mu_0(\rho)}{dx}
\end{equation}
where 
\begin{equation}
\mu_0(\rho)=\ln(\rho V_0^2(\rho)) 
\end{equation}

which acts an affinity driving the hydrodynamic current. 
By taking the first derivative of the CGF with respect to $\lambda$, the average current for small $\lambda$ is 
\begin{equation}
\langle J \rangle_{\lambda}=2\mathcal{D}(\rho)\lambda + \mathcal{O}(\lambda^3)
\end{equation}
so within the linear response regime, $\lambda$  is related to an affinity by
\begin{equation}
\lambda = -\frac{\rho}{2} \frac{d \mu_0(\rho)}{dx}
\end{equation}
and is the same result that could be obtained from the Gallavotti Cohen symmetry of the CGF. More generally, away from linear response, $\lambda$ and the affinity are still related, but this relation is more complicated~\cite{Andrieux2004}.

\section{Nonlinear hydrodynamic transport}
We can continue the Kramers Moyal expansion to the next non-zero term, which would include the fourth moment, $M^4=C^4_0-3(C^2_0)^2$. To this order, the hydrodynamic current would be given by
\begin{align}
J_{\rho}=-\mathcal{D}(\rho)\rho \left ( \frac{d \mu_0(\rho)}{dx} +\frac{d \mu_1(\rho)}{d x} \right )
 \end{align} 
 which includes an additional contribution to the effective chemical potential, $\mu_1(\rho)$ given by
  \begin{equation}
 \mu_1(\rho)=\alpha(\rho) \left (\frac{d \rho}{dx} \right )^2+\beta(\rho)\frac{d^2 \rho}{dx^2},  \tag{S18}
 \end{equation}
 with $\alpha(\rho)$ and $\beta(\rho)$ being coefficients that depend on the density, $v_0$, $D_t$ and $D_r$. Keeping the first order in the derivatives of the density we get
 \begin{align*}
\alpha(\rho) \mathcal{D}(\rho)\rho = &\, \frac{\left( -7+4D_r \right)}{4D_r^3}\left[V_0^3(\rho)\frac{\partial V_0}{\partial \rho}+\frac{3V_0^2(\rho)\rho}{2}\left |\frac{\partial V_0}{\partial \rho}\right|^2\right] \\ &\, +\frac{\partial V_0}{\partial \rho}\left[\frac{2V_0(\rho)}{Dr}
 +\frac{ \rho}{D_r}\frac{\partial V_0}{\partial \rho} \right]
 \end{align*}
 \begin{align*}
 \beta(\rho) \mathcal{D}(\rho)\rho =&\frac{1}{2}+\frac{\partial \mathcal{D}(\rho) }{\partial \rho }+\\
 &\frac{\left( -7+4D_r \right)}{8D_r^3}\left[ \frac{V_0^4(\rho)}{4} +\frac{\rho V_0^3(\rho)}{8}\frac{\partial V_0}{\partial \rho}\right]  
 \end{align*}
which shows that our framework allows us to quantify higher order response with the next order response containing  non-gradient terms that would give rise to interfacial energy-like terms necessary to stabilize two phases, $d^2\rho/dx^2$ and $|d \rho/dx|^2$~\cite{Cates2013,Cates2015,Speck2015}.


\begin{thebibliography}{98}%
\makeatletter
\providecommand \@ifxundefined [1]{%
 \@ifx{#1\undefined}
}%
\providecommand \@ifnum [1]{%
 \ifnum #1\expandafter \@firstoftwo
 \else \expandafter \@secondoftwo
 \fi
}%
\providecommand \@ifx [1]{%
 \ifx #1\expandafter \@firstoftwo
 \else \expandafter \@secondoftwo
 \fi
}%
\providecommand \natexlab [1]{#1}%
\providecommand \enquote  [1]{``#1''}%
\providecommand \bibnamefont  [1]{#1}%
\providecommand \bibfnamefont [1]{#1}%
\providecommand \citenamefont [1]{#1}%
\providecommand \href@noop [0]{\@secondoftwo}%
\providecommand \href [0]{\begingroup \@sanitize@url \@href}%
\providecommand \@href[1]{\@@startlink{#1}\@@href}%
\providecommand \@@href[1]{\endgroup#1\@@endlink}%
\providecommand \@sanitize@url [0]{\catcode `\\12\catcode `\$12\catcode
  `\&12\catcode `\#12\catcode `\^12\catcode `\_12\catcode `\%12\relax}%
\providecommand \@@startlink[1]{}%
\providecommand \@@endlink[0]{}%
\providecommand \url  [0]{\begingroup\@sanitize@url \@url }%
\providecommand \@url [1]{\endgroup\@href {#1}{\urlprefix }}%
\providecommand \urlprefix  [0]{URL }%
\providecommand \Eprint [0]{\href }%
\providecommand \doibase [0]{http://dx.doi.org/}%
\providecommand \selectlanguage [0]{\@gobble}%
\providecommand \bibinfo  [0]{\@secondoftwo}%
\providecommand \bibfield  [0]{\@secondoftwo}%
\providecommand \translation [1]{[#1]}%
\providecommand \BibitemOpen [0]{}%
\providecommand \bibitemStop [0]{}%
\providecommand \bibitemNoStop [0]{.\EOS\space}%
\providecommand \EOS [0]{\spacefactor3000\relax}%
\providecommand \BibitemShut  [1]{\csname bibitem#1\endcsname}%
\let\auto@bib@innerbib\@empty
\bibitem [{\citenamefont {Speck}\ and\ \citenamefont
  {Seifert}(2006)}]{Speck2006}%
  \BibitemOpen
  \bibfield  {author} {\bibinfo {author} {\bibfnamefont {T.}~\bibnamefont
  {Speck}}\ and\ \bibinfo {author} {\bibfnamefont {U.}~\bibnamefont
  {Seifert}},\ }\href@noop {} {\bibfield  {journal} {\bibinfo  {journal}
  {Europhys. Lett.}\ }\textbf {\bibinfo {volume} {74}},\ \bibinfo {pages} {391}
  (\bibinfo {year} {2006})}\BibitemShut {NoStop}%
\bibitem [{\citenamefont {Prost}\ \emph {et~al.}(2009)\citenamefont {Prost},
  \citenamefont {Joanny},\ and\ \citenamefont {Parrondo}}]{Prost2009}%
  \BibitemOpen
  \bibfield  {author} {\bibinfo {author} {\bibfnamefont {J.}~\bibnamefont
  {Prost}}, \bibinfo {author} {\bibfnamefont {J.~F.}\ \bibnamefont {Joanny}}, \
  and\ \bibinfo {author} {\bibfnamefont {J.~M.}\ \bibnamefont {Parrondo}},\
  }\href@noop {} {\bibfield  {journal} {\bibinfo  {journal} {Phys. Rev. Lett.}\
  }\textbf {\bibinfo {volume} {103}},\ \bibinfo {pages} {1} (\bibinfo {year}
  {2009})}\BibitemShut {NoStop}%
\bibitem [{\citenamefont {Baiesi}\ \emph {et~al.}(2009)\citenamefont {Baiesi},
  \citenamefont {Maes},\ and\ \citenamefont {Wynants}}]{Baiesi2009}%
  \BibitemOpen
  \bibfield  {author} {\bibinfo {author} {\bibfnamefont {M.}~\bibnamefont
  {Baiesi}}, \bibinfo {author} {\bibfnamefont {C.}~\bibnamefont {Maes}}, \ and\
  \bibinfo {author} {\bibfnamefont {B.}~\bibnamefont {Wynants}},\ }\href@noop
  {} {\bibfield  {journal} {\bibinfo  {journal} {Phys. Rev. Lett.}\ }\textbf
  {\bibinfo {volume} {103}},\ \bibinfo {pages} {1} (\bibinfo {year}
  {2009})}\BibitemShut {NoStop}%
\bibitem [{\citenamefont {Seifert}\ and\ \citenamefont
  {Speck}(2010)}]{Seifert2010}%
  \BibitemOpen
  \bibfield  {author} {\bibinfo {author} {\bibfnamefont {U.}~\bibnamefont
  {Seifert}}\ and\ \bibinfo {author} {\bibfnamefont {T.}~\bibnamefont
  {Speck}},\ }\href@noop {} {\bibfield  {journal} {\bibinfo  {journal}
  {Europhys. Lett.}\ }\textbf {\bibinfo {volume} {89}} (\bibinfo {year}
  {2010})}\BibitemShut {NoStop}%
\bibitem [{\citenamefont {Chetrite}\ and\ \citenamefont
  {Gupta}(2011)}]{Chetrite2011}%
  \BibitemOpen
  \bibfield  {author} {\bibinfo {author} {\bibfnamefont {R.}~\bibnamefont
  {Chetrite}}\ and\ \bibinfo {author} {\bibfnamefont {S.}~\bibnamefont
  {Gupta}},\ }\href@noop {} {\bibfield  {journal} {\bibinfo  {journal} {J.
  Stat. Phys.}\ }\textbf {\bibinfo {volume} {143}},\ \bibinfo {pages} {543}
  (\bibinfo {year} {2011})}\BibitemShut {NoStop}%
\bibitem [{\citenamefont {Baiesi}\ and\ \citenamefont
  {Maes}(2013)}]{Baiesi2013}%
  \BibitemOpen
  \bibfield  {author} {\bibinfo {author} {\bibfnamefont {M.}~\bibnamefont
  {Baiesi}}\ and\ \bibinfo {author} {\bibfnamefont {C.}~\bibnamefont {Maes}},\
  }\href@noop {} {\bibfield  {journal} {\bibinfo  {journal} {New J. Phys.}\
  }\textbf {\bibinfo {volume} {15}} (\bibinfo {year} {2013})}\BibitemShut
  {NoStop}%
\bibitem [{\citenamefont {Maes}(2014)}]{Maes2014}%
  \BibitemOpen
  \bibfield  {author} {\bibinfo {author} {\bibfnamefont {C.}~\bibnamefont
  {Maes}},\ }\href@noop {} {\bibfield  {journal} {\bibinfo  {journal} {J. Stat.
  Phys.}\ }\textbf {\bibinfo {volume} {154}},\ \bibinfo {pages} {705} (\bibinfo
  {year} {2014})}\BibitemShut {NoStop}%
\bibitem [{\citenamefont {Evans}\ and\ \citenamefont
  {Searles}(1994)}]{Evans1994}%
  \BibitemOpen
  \bibfield  {author} {\bibinfo {author} {\bibfnamefont {D.~J.}\ \bibnamefont
  {Evans}}\ and\ \bibinfo {author} {\bibfnamefont {D.~J.}\ \bibnamefont
  {Searles}},\ }\href@noop {} {\bibfield  {journal} {\bibinfo  {journal} {Phys.
  Rev. E}\ }\textbf {\bibinfo {volume} {50}},\ \bibinfo {pages} {1645}
  (\bibinfo {year} {1994})}\BibitemShut {NoStop}%
\bibitem [{\citenamefont {Kurchan}(1997)}]{Kurchan1997}%
  \BibitemOpen
  \bibfield  {author} {\bibinfo {author} {\bibfnamefont {J.}~\bibnamefont
  {Kurchan}},\ }\href
  {http://arxiv.org/abs/cond-mat/9709304%0Ahttp://dx.doi.org/10.1088/0305-4470/31/16/003}
  {\bibfield  {journal} {\bibinfo  {journal} {Phys. Rev. E}\ } (\bibinfo {year}
  {1997})}\BibitemShut {NoStop}%
\bibitem [{\citenamefont {Crooks}(1999)}]{Crooks1999}%
  \BibitemOpen
  \bibfield  {author} {\bibinfo {author} {\bibfnamefont {G.~E.}\ \bibnamefont
  {Crooks}},\ }\href@noop {} {\bibfield  {journal} {\bibinfo  {journal} {Phys.
  Rev. E}\ }\textbf {\bibinfo {volume} {60}},\ \bibinfo {pages} {2721}
  (\bibinfo {year} {1999})}\BibitemShut {NoStop}%
\bibitem [{\citenamefont {Seifert}(2005)}]{Seifert2005}%
  \BibitemOpen
  \bibfield  {author} {\bibinfo {author} {\bibfnamefont {U.}~\bibnamefont
  {Seifert}},\ }\href@noop {} {\bibfield  {journal} {\bibinfo  {journal} {Phys.
  Rev. Lett.}\ }\textbf {\bibinfo {volume} {95}},\ \bibinfo {pages} {1}
  (\bibinfo {year} {2005})}\BibitemShut {NoStop}%
\bibitem [{\citenamefont {Jarzynski}(1997)}]{Jarzynski1997}%
  \BibitemOpen
  \bibfield  {author} {\bibinfo {author} {\bibfnamefont {C.}~\bibnamefont
  {Jarzynski}},\ }\href@noop {} {\bibfield  {journal} {\bibinfo  {journal}
  {Phys. Rev. Lett.}\ }\textbf {\bibinfo {volume} {78}},\ \bibinfo {pages}
  {2690} (\bibinfo {year} {1997})}\BibitemShut {NoStop}%
\bibitem [{\citenamefont {Gallavotti}\ and\ \citenamefont
  {Cohen}(1995)}]{Gallavotti1995}%
  \BibitemOpen
  \bibfield  {author} {\bibinfo {author} {\bibfnamefont {G.}~\bibnamefont
  {Gallavotti}}\ and\ \bibinfo {author} {\bibfnamefont {E.~G.}\ \bibnamefont
  {Cohen}},\ }\href@noop {} {\bibfield  {journal} {\bibinfo  {journal} {Phys.
  Rev. Lett.}\ }\textbf {\bibinfo {volume} {74}},\ \bibinfo {pages} {2694}
  (\bibinfo {year} {1995})}\BibitemShut {NoStop}%
\bibitem [{\citenamefont {Lebowitz}\ and\ \citenamefont
  {Spohn}(1999)}]{Lebowitz1998b}%
  \BibitemOpen
  \bibfield  {author} {\bibinfo {author} {\bibfnamefont {J.~L.}\ \bibnamefont
  {Lebowitz}}\ and\ \bibinfo {author} {\bibfnamefont {H.}~\bibnamefont
  {Spohn}},\ }\href
  {http://arxiv.org/abs/cond-mat/9811220%0Ahttp://dx.doi.org/10.1023/A:1004589714161}
  {\bibfield  {journal} {\bibinfo  {journal} {Springer}\ ,\ \bibinfo {pages}
  {333}} (\bibinfo {year} {1999})}\BibitemShut {NoStop}%
\bibitem [{\citenamefont {Barato}\ and\ \citenamefont
  {Seifert}(2015)}]{Barato2015}%
  \BibitemOpen
  \bibfield  {author} {\bibinfo {author} {\bibfnamefont {A.~C.}\ \bibnamefont
  {Barato}}\ and\ \bibinfo {author} {\bibfnamefont {U.}~\bibnamefont
  {Seifert}},\ }\href@noop {} {\bibfield  {journal} {\bibinfo  {journal} {Phys.
  Rev. Lett.}\ }\textbf {\bibinfo {volume} {114}},\ \bibinfo {pages} {1}
  (\bibinfo {year} {2015})}\BibitemShut {NoStop}%
\bibitem [{\citenamefont {Gingrich}\ \emph {et~al.}(2016)\citenamefont
  {Gingrich}, \citenamefont {Horowitz}, \citenamefont {Perunov},\ and\
  \citenamefont {England}}]{Gingrich2016}%
  \BibitemOpen
  \bibfield  {author} {\bibinfo {author} {\bibfnamefont {T.~R.}\ \bibnamefont
  {Gingrich}}, \bibinfo {author} {\bibfnamefont {J.~M.}\ \bibnamefont
  {Horowitz}}, \bibinfo {author} {\bibfnamefont {N.}~\bibnamefont {Perunov}}, \
  and\ \bibinfo {author} {\bibfnamefont {J.~L.}\ \bibnamefont {England}},\
  }\href@noop {} {\bibfield  {journal} {\bibinfo  {journal} {Phys. Rev. Lett.}\
  }\textbf {\bibinfo {volume} {116}},\ \bibinfo {pages} {1} (\bibinfo {year}
  {2016})}\BibitemShut {NoStop}%
\bibitem [{\citenamefont {Polettini}\ \emph {et~al.}(2016)\citenamefont
  {Polettini}, \citenamefont {Lazarescu},\ and\ \citenamefont
  {Esposito}}]{Polettini2016}%
  \BibitemOpen
  \bibfield  {author} {\bibinfo {author} {\bibfnamefont {M.}~\bibnamefont
  {Polettini}}, \bibinfo {author} {\bibfnamefont {A.}~\bibnamefont
  {Lazarescu}}, \ and\ \bibinfo {author} {\bibfnamefont {M.}~\bibnamefont
  {Esposito}},\ }\href@noop {} {\bibfield  {journal} {\bibinfo  {journal}
  {Phys. Rev. E}\ }\textbf {\bibinfo {volume} {94}},\ \bibinfo {pages} {1}
  (\bibinfo {year} {2016})}\BibitemShut {NoStop}%
\bibitem [{\citenamefont {Chetrite}\ and\ \citenamefont
  {Touchette}(2015)}]{Chetrite2015}%
  \BibitemOpen
  \bibfield  {author} {\bibinfo {author} {\bibfnamefont {R.}~\bibnamefont
  {Chetrite}}\ and\ \bibinfo {author} {\bibfnamefont {H.}~\bibnamefont
  {Touchette}},\ }\href@noop {} {\bibfield  {journal} {\bibinfo  {journal}
  {Ann. Hen. P.}\ }\textbf {\bibinfo {volume} {16}},\ \bibinfo {pages} {2005}
  (\bibinfo {year} {2015})}\BibitemShut {NoStop}%
\bibitem [{\citenamefont {Lebowitz}\ and\ \citenamefont
  {Spohn}(1998)}]{Lebowitz1998}%
  \BibitemOpen
  \bibfield  {author} {\bibinfo {author} {\bibfnamefont {J.~L.}\ \bibnamefont
  {Lebowitz}}\ and\ \bibinfo {author} {\bibfnamefont {H.}~\bibnamefont
  {Spohn}},\ }\href
  {http://arxiv.org/abs/cond-mat/9811220%0Ahttp://dx.doi.org/10.1023/A:1004589714161}
  {\bibfield  {journal} {\bibinfo  {journal} {Phys. Rev. E}\ ,\ \bibinfo
  {pages} {333}} (\bibinfo {year} {1998})}\BibitemShut {NoStop}%
\bibitem [{\citenamefont {Speck}(2016)}]{Speck2016}%
  \BibitemOpen
  \bibfield  {author} {\bibinfo {author} {\bibfnamefont {T.}~\bibnamefont
  {Speck}},\ }\href@noop {} {\bibfield  {journal} {\bibinfo  {journal} {Phys.
  Rev. E}\ }\textbf {\bibinfo {volume} {94}},\ \bibinfo {pages} {1} (\bibinfo
  {year} {2016})}\BibitemShut {NoStop}%
\bibitem [{\citenamefont {Gao}\ and\ \citenamefont {Limmer}(2017)}]{Gao2017}%
  \BibitemOpen
  \bibfield  {author} {\bibinfo {author} {\bibfnamefont {C.~Y.}\ \bibnamefont
  {Gao}}\ and\ \bibinfo {author} {\bibfnamefont {D.~T.}\ \bibnamefont
  {Limmer}},\ }\href@noop {} {\bibfield  {journal} {\bibinfo  {journal}
  {Entropy}\ }\textbf {\bibinfo {volume} {19}},\ \bibinfo {pages} {1} (\bibinfo
  {year} {2017})}\BibitemShut {NoStop}%
\bibitem [{\citenamefont {Gaspard}\ and\ \citenamefont
  {Kapral}(2017)}]{Gaspard2017}%
  \BibitemOpen
  \bibfield  {author} {\bibinfo {author} {\bibfnamefont {P.}~\bibnamefont
  {Gaspard}}\ and\ \bibinfo {author} {\bibfnamefont {R.}~\bibnamefont
  {Kapral}},\ }\href@noop {} {\bibfield  {journal} {\bibinfo  {journal} {J.
  Chem. Phys.}\ }\textbf {\bibinfo {volume} {147}} (\bibinfo {year}
  {2017})}\BibitemShut {NoStop}%
\bibitem [{\citenamefont {Cates}\ and\ \citenamefont
  {Tailleur}(2015)}]{Cates2015}%
  \BibitemOpen
  \bibfield  {author} {\bibinfo {author} {\bibfnamefont {M.~E.}\ \bibnamefont
  {Cates}}\ and\ \bibinfo {author} {\bibfnamefont {J.}~\bibnamefont
  {Tailleur}},\ }\href
  {http://www.annualreviews.org/doi/10.1146/annurev-conmatphys-031214-014710}
  {\bibfield  {journal} {\bibinfo  {journal} {Ann. Rev. Cond. Mat. Phys.}\
  }\textbf {\bibinfo {volume} {6}},\ \bibinfo {pages} {219} (\bibinfo {year}
  {2015})}\BibitemShut {NoStop}%
\bibitem [{\citenamefont {Redner}\ \emph {et~al.}(2013)\citenamefont {Redner},
  \citenamefont {Hagan},\ and\ \citenamefont {Baskaran}}]{Redner2013}%
  \BibitemOpen
  \bibfield  {author} {\bibinfo {author} {\bibfnamefont {G.~S.}\ \bibnamefont
  {Redner}}, \bibinfo {author} {\bibfnamefont {M.~F.}\ \bibnamefont {Hagan}}, \
  and\ \bibinfo {author} {\bibfnamefont {A.}~\bibnamefont {Baskaran}},\
  }\href@noop {} {\bibfield  {journal} {\bibinfo  {journal} {Phys. Rev. Lett.}\
  }\textbf {\bibinfo {volume} {110}},\ \bibinfo {pages} {1} (\bibinfo {year}
  {2013})}\BibitemShut {NoStop}%
\bibitem [{\citenamefont {Schaller}\ \emph {et~al.}(2010)\citenamefont
  {Schaller}, \citenamefont {Weber}, \citenamefont {Semmrich}, \citenamefont
  {Frey},\ and\ \citenamefont {Bausch}}]{Schaller2010}%
  \BibitemOpen
  \bibfield  {author} {\bibinfo {author} {\bibfnamefont {V.}~\bibnamefont
  {Schaller}}, \bibinfo {author} {\bibfnamefont {C.}~\bibnamefont {Weber}},
  \bibinfo {author} {\bibfnamefont {C.}~\bibnamefont {Semmrich}}, \bibinfo
  {author} {\bibfnamefont {E.}~\bibnamefont {Frey}}, \ and\ \bibinfo {author}
  {\bibfnamefont {A.~R.}\ \bibnamefont {Bausch}},\ }\href
  {http://dx.doi.org/10.1038/nature09312} {\bibfield  {journal} {\bibinfo
  {journal} {Nature}\ }\textbf {\bibinfo {volume} {467}},\ \bibinfo {pages}
  {73} (\bibinfo {year} {2010})}\BibitemShut {NoStop}%
\bibitem [{\citenamefont {Sumino}\ \emph {et~al.}(2012)\citenamefont {Sumino},
  \citenamefont {Nagai}, \citenamefont {Shitaka}, \citenamefont {Tanaka},
  \citenamefont {Yoshikawa}, \citenamefont {Chat{\'{e}}},\ and\ \citenamefont
  {Oiwa}}]{Sumino2012}%
  \BibitemOpen
  \bibfield  {author} {\bibinfo {author} {\bibfnamefont {Y.}~\bibnamefont
  {Sumino}}, \bibinfo {author} {\bibfnamefont {K.~H.}\ \bibnamefont {Nagai}},
  \bibinfo {author} {\bibfnamefont {Y.}~\bibnamefont {Shitaka}}, \bibinfo
  {author} {\bibfnamefont {D.}~\bibnamefont {Tanaka}}, \bibinfo {author}
  {\bibfnamefont {K.}~\bibnamefont {Yoshikawa}}, \bibinfo {author}
  {\bibfnamefont {H.}~\bibnamefont {Chat{\'{e}}}}, \ and\ \bibinfo {author}
  {\bibfnamefont {K.}~\bibnamefont {Oiwa}},\ }\href@noop {} {\bibfield
  {journal} {\bibinfo  {journal} {Nature}\ }\textbf {\bibinfo {volume} {483}},\
  \bibinfo {pages} {448} (\bibinfo {year} {2012})}\BibitemShut {NoStop}%
\bibitem [{\citenamefont {Schaller}\ \emph {et~al.}(2011)\citenamefont
  {Schaller}, \citenamefont {Weber}, \citenamefont {Hammerich}, \citenamefont
  {Frey},\ and\ \citenamefont {Bausch}}]{Schaller2011}%
  \BibitemOpen
  \bibfield  {author} {\bibinfo {author} {\bibfnamefont {V.}~\bibnamefont
  {Schaller}}, \bibinfo {author} {\bibfnamefont {C.~A.}\ \bibnamefont {Weber}},
  \bibinfo {author} {\bibfnamefont {B.}~\bibnamefont {Hammerich}}, \bibinfo
  {author} {\bibfnamefont {E.}~\bibnamefont {Frey}}, \ and\ \bibinfo {author}
  {\bibfnamefont {A.~R.}\ \bibnamefont {Bausch}},\ }\href
  {http://www.pnas.org/cgi/doi/10.1073/pnas.1107540108} {\bibfield  {journal}
  {\bibinfo  {journal} {Proc. Natl. Acad. Sci. U.S.A.}\ }\textbf {\bibinfo
  {volume} {108}},\ \bibinfo {pages} {19183} (\bibinfo {year}
  {2011})}\BibitemShut {NoStop}%
\bibitem [{\citenamefont {Berg}\ and\ \citenamefont {Brown}(1972)}]{Berg1972}%
  \BibitemOpen
  \bibfield  {author} {\bibinfo {author} {\bibfnamefont {H.~C.}\ \bibnamefont
  {Berg}}\ and\ \bibinfo {author} {\bibfnamefont {D.~A.}\ \bibnamefont
  {Brown}},\ }\href@noop {} {\bibfield  {journal} {\bibinfo  {journal}
  {Nature}\ }\textbf {\bibinfo {volume} {239}},\ \bibinfo {pages} {500}
  (\bibinfo {year} {1972})}\BibitemShut {NoStop}%
\bibitem [{\citenamefont {Shapiro}(1995)}]{Shapiro1995}%
  \BibitemOpen
  \bibfield  {author} {\bibinfo {author} {\bibfnamefont {J.~A.}\ \bibnamefont
  {Shapiro}},\ }\href@noop {} {\bibfield  {journal} {\bibinfo  {journal}
  {BioEssays}\ }\textbf {\bibinfo {volume} {17}},\ \bibinfo {pages} {597}
  (\bibinfo {year} {1995})}\BibitemShut {NoStop}%
\bibitem [{\citenamefont {Dombrowski}\ \emph {et~al.}(2004)\citenamefont
  {Dombrowski}, \citenamefont {Cisneros}, \citenamefont {Chatkaew},
  \citenamefont {Goldstein},\ and\ \citenamefont {Kessler}}]{Dombrowski2004}%
  \BibitemOpen
  \bibfield  {author} {\bibinfo {author} {\bibfnamefont {C.}~\bibnamefont
  {Dombrowski}}, \bibinfo {author} {\bibfnamefont {L.}~\bibnamefont
  {Cisneros}}, \bibinfo {author} {\bibfnamefont {S.}~\bibnamefont {Chatkaew}},
  \bibinfo {author} {\bibfnamefont {R.~E.}\ \bibnamefont {Goldstein}}, \ and\
  \bibinfo {author} {\bibfnamefont {J.~O.}\ \bibnamefont {Kessler}},\
  }\href@noop {} {\bibfield  {journal} {\bibinfo  {journal} {Phys. Rev. Lett.}\
  }\textbf {\bibinfo {volume} {93}},\ \bibinfo {pages} {2} (\bibinfo {year}
  {2004})}\BibitemShut {NoStop}%
\bibitem [{\citenamefont {Hill}\ \emph {et~al.}(2007)\citenamefont {Hill},
  \citenamefont {Kalkanci}, \citenamefont {McMurry},\ and\ \citenamefont
  {Koser}}]{Hill2007}%
  \BibitemOpen
  \bibfield  {author} {\bibinfo {author} {\bibfnamefont {J.}~\bibnamefont
  {Hill}}, \bibinfo {author} {\bibfnamefont {O.}~\bibnamefont {Kalkanci}},
  \bibinfo {author} {\bibfnamefont {J.~L.}\ \bibnamefont {McMurry}}, \ and\
  \bibinfo {author} {\bibfnamefont {H.}~\bibnamefont {Koser}},\ }\href@noop {}
  {\bibfield  {journal} {\bibinfo  {journal} {Phys. Rev. Lett.}\ }\textbf
  {\bibinfo {volume} {98}},\ \bibinfo {pages} {1} (\bibinfo {year}
  {2007})}\BibitemShut {NoStop}%
\bibitem [{\citenamefont {Lauga}\ \emph {et~al.}(2006)\citenamefont {Lauga},
  \citenamefont {DiLuzio}, \citenamefont {Whitesides},\ and\ \citenamefont
  {Stone}}]{Lauga2006}%
  \BibitemOpen
  \bibfield  {author} {\bibinfo {author} {\bibfnamefont {E.}~\bibnamefont
  {Lauga}}, \bibinfo {author} {\bibfnamefont {W.~R.}\ \bibnamefont {DiLuzio}},
  \bibinfo {author} {\bibfnamefont {G.~M.}\ \bibnamefont {Whitesides}}, \ and\
  \bibinfo {author} {\bibfnamefont {H.~A.}\ \bibnamefont {Stone}},\ }\href
  {http://dx.doi.org/10.1529/biophysj.105.069401} {\bibfield  {journal}
  {\bibinfo  {journal} {Biophys. J.}\ }\textbf {\bibinfo {volume} {90}},\
  \bibinfo {pages} {400} (\bibinfo {year} {2006})}\BibitemShut {NoStop}%
\bibitem [{\citenamefont {Fu}\ \emph {et~al.}(2012)\citenamefont {Fu},
  \citenamefont {Tang}, \citenamefont {Liu}, \citenamefont {Huang},
  \citenamefont {Hwa},\ and\ \citenamefont {Lenz}}]{Fu2012}%
  \BibitemOpen
  \bibfield  {author} {\bibinfo {author} {\bibfnamefont {X.}~\bibnamefont
  {Fu}}, \bibinfo {author} {\bibfnamefont {L.~H.}\ \bibnamefont {Tang}},
  \bibinfo {author} {\bibfnamefont {C.}~\bibnamefont {Liu}}, \bibinfo {author}
  {\bibfnamefont {J.~D.}\ \bibnamefont {Huang}}, \bibinfo {author}
  {\bibfnamefont {T.}~\bibnamefont {Hwa}}, \ and\ \bibinfo {author}
  {\bibfnamefont {P.}~\bibnamefont {Lenz}},\ }\href@noop {} {\bibfield
  {journal} {\bibinfo  {journal} {Phys. Rev. Lett.}\ }\textbf {\bibinfo
  {volume} {108}},\ \bibinfo {pages} {1} (\bibinfo {year} {2012})}\BibitemShut
  {NoStop}%
\bibitem [{\citenamefont {Parsek}\ and\ \citenamefont
  {Greenberg}(2005)}]{Parsek2005}%
  \BibitemOpen
  \bibfield  {author} {\bibinfo {author} {\bibfnamefont {M.~R.}\ \bibnamefont
  {Parsek}}\ and\ \bibinfo {author} {\bibfnamefont {E.~P.}\ \bibnamefont
  {Greenberg}},\ }\href@noop {} {\bibfield  {journal} {\bibinfo  {journal}
  {Trends Microbiol.}\ }\textbf {\bibinfo {volume} {13}},\ \bibinfo {pages}
  {27} (\bibinfo {year} {2005})}\BibitemShut {NoStop}%
\bibitem [{\citenamefont {Palacci}\ \emph {et~al.}(2013)\citenamefont
  {Palacci}, \citenamefont {Sacanna}, \citenamefont {Steinberg}, \citenamefont
  {Pine},\ and\ \citenamefont {Chaikin}}]{Palacci2013}%
  \BibitemOpen
  \bibfield  {author} {\bibinfo {author} {\bibfnamefont {J.}~\bibnamefont
  {Palacci}}, \bibinfo {author} {\bibfnamefont {S.}~\bibnamefont {Sacanna}},
  \bibinfo {author} {\bibfnamefont {A.~P.}\ \bibnamefont {Steinberg}}, \bibinfo
  {author} {\bibfnamefont {D.~J.}\ \bibnamefont {Pine}}, \ and\ \bibinfo
  {author} {\bibfnamefont {P.~M.}\ \bibnamefont {Chaikin}},\ }\href
  {http://www.sciencemag.org/content/339/6122/936.full.pdf} {\bibfield
  {journal} {\bibinfo  {journal} {Science}\ }\textbf {\bibinfo {volume}
  {339}},\ \bibinfo {pages} {936} (\bibinfo {year} {2013})}\BibitemShut
  {NoStop}%
\bibitem [{\citenamefont {Narayan}\ \emph {et~al.}(2006)\citenamefont
  {Narayan}, \citenamefont {Menon},\ and\ \citenamefont
  {Ramaswamy}}]{Narayan2006}%
  \BibitemOpen
  \bibfield  {author} {\bibinfo {author} {\bibfnamefont {V.}~\bibnamefont
  {Narayan}}, \bibinfo {author} {\bibfnamefont {N.}~\bibnamefont {Menon}}, \
  and\ \bibinfo {author} {\bibfnamefont {S.}~\bibnamefont {Ramaswamy}},\
  }\href@noop {} {\bibfield  {journal} {\bibinfo  {journal} {J. Stat. Mech.}\ }
  (\bibinfo {year} {2006})}\BibitemShut {NoStop}%
\bibitem [{\citenamefont {Howse}\ \emph {et~al.}(2007)\citenamefont {Howse},
  \citenamefont {Jones}, \citenamefont {Ryan}, \citenamefont {Gough},
  \citenamefont {Vafabakhsh},\ and\ \citenamefont {Golestanian}}]{Howse2007}%
  \BibitemOpen
  \bibfield  {author} {\bibinfo {author} {\bibfnamefont {J.~R.}\ \bibnamefont
  {Howse}}, \bibinfo {author} {\bibfnamefont {R.~A.}\ \bibnamefont {Jones}},
  \bibinfo {author} {\bibfnamefont {A.~J.}\ \bibnamefont {Ryan}}, \bibinfo
  {author} {\bibfnamefont {T.}~\bibnamefont {Gough}}, \bibinfo {author}
  {\bibfnamefont {R.}~\bibnamefont {Vafabakhsh}}, \ and\ \bibinfo {author}
  {\bibfnamefont {R.}~\bibnamefont {Golestanian}},\ }\href@noop {} {\bibfield
  {journal} {\bibinfo  {journal} {Phys. Rev. Lett.}\ }\textbf {\bibinfo
  {volume} {99}},\ \bibinfo {pages} {8} (\bibinfo {year} {2007})}\BibitemShut
  {NoStop}%
\bibitem [{\citenamefont {Walther}\ and\ \citenamefont
  {M{\"{u}}ller}(2008)}]{Walther2008}%
  \BibitemOpen
  \bibfield  {author} {\bibinfo {author} {\bibfnamefont {A.}~\bibnamefont
  {Walther}}\ and\ \bibinfo {author} {\bibfnamefont {A.~H.~E.}\ \bibnamefont
  {M{\"{u}}ller}},\ }\href {http://xlink.rsc.org/?DOI=b718131k} {\bibfield
  {journal} {\bibinfo  {journal} {Soft Matter}\ }\textbf {\bibinfo {volume}
  {4}},\ \bibinfo {pages} {663} (\bibinfo {year} {2008})}\BibitemShut {NoStop}%
\bibitem [{\citenamefont {Bricard}\ \emph {et~al.}(2013)\citenamefont
  {Bricard}, \citenamefont {Caussin}, \citenamefont {Desreumaux}, \citenamefont
  {Dauchot},\ and\ \citenamefont {Bartolo}}]{Bricard2013}%
  \BibitemOpen
  \bibfield  {author} {\bibinfo {author} {\bibfnamefont {A.}~\bibnamefont
  {Bricard}}, \bibinfo {author} {\bibfnamefont {J.~B.}\ \bibnamefont
  {Caussin}}, \bibinfo {author} {\bibfnamefont {N.}~\bibnamefont {Desreumaux}},
  \bibinfo {author} {\bibfnamefont {O.}~\bibnamefont {Dauchot}}, \ and\
  \bibinfo {author} {\bibfnamefont {D.}~\bibnamefont {Bartolo}},\ }\href
  {http://dx.doi.org/10.1038/nature12673} {\bibfield  {journal} {\bibinfo
  {journal} {Nature}\ }\textbf {\bibinfo {volume} {503}},\ \bibinfo {pages}
  {95} (\bibinfo {year} {2013})}\BibitemShut {NoStop}%
\bibitem [{\citenamefont {Bechinger}\ \emph {et~al.}(2016)\citenamefont
  {Bechinger}, \citenamefont {Di~Leonardo}, \citenamefont {L{\"{o}}wen},
  \citenamefont {Reichhardt}, \citenamefont {Volpe},\ and\ \citenamefont
  {Volpe}}]{Bechinger2016}%
  \BibitemOpen
  \bibfield  {author} {\bibinfo {author} {\bibfnamefont {C.}~\bibnamefont
  {Bechinger}}, \bibinfo {author} {\bibfnamefont {R.}~\bibnamefont
  {Di~Leonardo}}, \bibinfo {author} {\bibfnamefont {H.}~\bibnamefont
  {L{\"{o}}wen}}, \bibinfo {author} {\bibfnamefont {C.}~\bibnamefont
  {Reichhardt}}, \bibinfo {author} {\bibfnamefont {G.}~\bibnamefont {Volpe}}, \
  and\ \bibinfo {author} {\bibfnamefont {G.}~\bibnamefont {Volpe}},\
  }\href@noop {} {\bibfield  {journal} {\bibinfo  {journal} {Rev. Mod. Phys}\
  }\textbf {\bibinfo {volume} {88}} (\bibinfo {year} {2016})}\BibitemShut
  {NoStop}%
\bibitem [{\citenamefont {Buttinoni}\ \emph {et~al.}(2013)\citenamefont
  {Buttinoni}, \citenamefont {Bialk{\'{e}}}, \citenamefont {K{\"{u}}mmel},
  \citenamefont {L{\"{o}}wen}, \citenamefont {Bechinger},\ and\ \citenamefont
  {Speck}}]{Buttinoni2013}%
  \BibitemOpen
  \bibfield  {author} {\bibinfo {author} {\bibfnamefont {I.}~\bibnamefont
  {Buttinoni}}, \bibinfo {author} {\bibfnamefont {J.}~\bibnamefont
  {Bialk{\'{e}}}}, \bibinfo {author} {\bibfnamefont {F.}~\bibnamefont
  {K{\"{u}}mmel}}, \bibinfo {author} {\bibfnamefont {H.}~\bibnamefont
  {L{\"{o}}wen}}, \bibinfo {author} {\bibfnamefont {C.}~\bibnamefont
  {Bechinger}}, \ and\ \bibinfo {author} {\bibfnamefont {T.}~\bibnamefont
  {Speck}},\ }\href@noop {} {\bibfield  {journal} {\bibinfo  {journal} {Phys.
  Rev. Lett.}\ }\textbf {\bibinfo {volume} {110}},\ \bibinfo {pages} {1}
  (\bibinfo {year} {2013})}\BibitemShut {NoStop}%
\bibitem [{\citenamefont {Elgeti}\ \emph {et~al.}(2015)\citenamefont {Elgeti},
  \citenamefont {Winkler},\ and\ \citenamefont {Gompper}}]{Elgeti2015}%
  \BibitemOpen
  \bibfield  {author} {\bibinfo {author} {\bibfnamefont {J.}~\bibnamefont
  {Elgeti}}, \bibinfo {author} {\bibfnamefont {R.~G.}\ \bibnamefont {Winkler}},
  \ and\ \bibinfo {author} {\bibfnamefont {G.}~\bibnamefont {Gompper}},\
  }\href@noop {} {\bibfield  {journal} {\bibinfo  {journal} {Rep. Prog. Phys}\
  }\textbf {\bibinfo {volume} {78}} (\bibinfo {year} {2015})}\BibitemShut
  {NoStop}%
\bibitem [{\citenamefont {Yang}\ \emph {et~al.}(2010)\citenamefont {Yang},
  \citenamefont {Marceau},\ and\ \citenamefont {Gompper}}]{Yang2010}%
  \BibitemOpen
  \bibfield  {author} {\bibinfo {author} {\bibfnamefont {Y.}~\bibnamefont
  {Yang}}, \bibinfo {author} {\bibfnamefont {V.}~\bibnamefont {Marceau}}, \
  and\ \bibinfo {author} {\bibfnamefont {G.}~\bibnamefont {Gompper}},\
  }\href@noop {} {\bibfield  {journal} {\bibinfo  {journal} {Phys. Rev. E}\
  }\textbf {\bibinfo {volume} {82}},\ \bibinfo {pages} {1} (\bibinfo {year}
  {2010})}\BibitemShut {NoStop}%
\bibitem [{\citenamefont {Peruani}(2016)}]{Peruani2016}%
  \BibitemOpen
  \bibfield  {author} {\bibinfo {author} {\bibfnamefont {F.}~\bibnamefont
  {Peruani}},\ }\href@noop {} {\bibfield  {journal} {\bibinfo  {journal} {Eur.
  Phys. J.}\ }\textbf {\bibinfo {volume} {225}},\ \bibinfo {pages} {2301}
  (\bibinfo {year} {2016})}\BibitemShut {NoStop}%
\bibitem [{\citenamefont {Ginelli}\ \emph {et~al.}(2010)\citenamefont
  {Ginelli}, \citenamefont {Peruani}, \citenamefont {B{\"{a}}r},\ and\
  \citenamefont {Chat{\'{e}}}}]{Ginelli2010}%
  \BibitemOpen
  \bibfield  {author} {\bibinfo {author} {\bibfnamefont {F.}~\bibnamefont
  {Ginelli}}, \bibinfo {author} {\bibfnamefont {F.}~\bibnamefont {Peruani}},
  \bibinfo {author} {\bibfnamefont {M.}~\bibnamefont {B{\"{a}}r}}, \ and\
  \bibinfo {author} {\bibfnamefont {H.}~\bibnamefont {Chat{\'{e}}}},\
  }\href@noop {} {\bibfield  {journal} {\bibinfo  {journal} {Phys. Rev. Lett.}\
  }\textbf {\bibinfo {volume} {104}},\ \bibinfo {pages} {1} (\bibinfo {year}
  {2010})}\BibitemShut {NoStop}%
\bibitem [{\citenamefont {Abkenar}\ \emph {et~al.}(2013)\citenamefont
  {Abkenar}, \citenamefont {Marx}, \citenamefont {Auth},\ and\ \citenamefont
  {Gompper}}]{Abkenar2013}%
  \BibitemOpen
  \bibfield  {author} {\bibinfo {author} {\bibfnamefont {M.}~\bibnamefont
  {Abkenar}}, \bibinfo {author} {\bibfnamefont {K.}~\bibnamefont {Marx}},
  \bibinfo {author} {\bibfnamefont {T.}~\bibnamefont {Auth}}, \ and\ \bibinfo
  {author} {\bibfnamefont {G.}~\bibnamefont {Gompper}},\ }\href@noop {}
  {\bibfield  {journal} {\bibinfo  {journal} {Phys. Rev. E}\ }\textbf {\bibinfo
  {volume} {88}},\ \bibinfo {pages} {1} (\bibinfo {year} {2013})}\BibitemShut
  {NoStop}%
\bibitem [{\citenamefont {Peruani}\ \emph {et~al.}(2012)\citenamefont
  {Peruani}, \citenamefont {Starru{\ss}}, \citenamefont {Jakovljevic},
  \citenamefont {S{\o}gaard-Andersen}, \citenamefont {Deutsch},\ and\
  \citenamefont {B{\"{a}}r}}]{Peruani2012}%
  \BibitemOpen
  \bibfield  {author} {\bibinfo {author} {\bibfnamefont {F.}~\bibnamefont
  {Peruani}}, \bibinfo {author} {\bibfnamefont {J.}~\bibnamefont
  {Starru{\ss}}}, \bibinfo {author} {\bibfnamefont {V.}~\bibnamefont
  {Jakovljevic}}, \bibinfo {author} {\bibfnamefont {L.}~\bibnamefont
  {S{\o}gaard-Andersen}}, \bibinfo {author} {\bibfnamefont {A.}~\bibnamefont
  {Deutsch}}, \ and\ \bibinfo {author} {\bibfnamefont {M.}~\bibnamefont
  {B{\"{a}}r}},\ }\href@noop {} {\bibfield  {journal} {\bibinfo  {journal}
  {Phys. Rev. Lett.}\ }\textbf {\bibinfo {volume} {108}},\ \bibinfo {pages} {1}
  (\bibinfo {year} {2012})}\BibitemShut {NoStop}%
\bibitem [{\citenamefont {Harvey}\ \emph {et~al.}(2013)\citenamefont {Harvey},
  \citenamefont {Alber}, \citenamefont {Tsimring},\ and\ \citenamefont
  {Aranson}}]{Harvey2013}%
  \BibitemOpen
  \bibfield  {author} {\bibinfo {author} {\bibfnamefont {C.~W.}\ \bibnamefont
  {Harvey}}, \bibinfo {author} {\bibfnamefont {M.}~\bibnamefont {Alber}},
  \bibinfo {author} {\bibfnamefont {L.~S.}\ \bibnamefont {Tsimring}}, \ and\
  \bibinfo {author} {\bibfnamefont {I.~S.}\ \bibnamefont {Aranson}},\
  }\href@noop {} {\bibfield  {journal} {\bibinfo  {journal} {New J. Phys.}\
  }\textbf {\bibinfo {volume} {15}} (\bibinfo {year} {2013})}\BibitemShut
  {NoStop}%
\bibitem [{\citenamefont {Chelakkot}\ \emph {et~al.}(2014)\citenamefont
  {Chelakkot}, \citenamefont {Gopinath}, \citenamefont {Mahadevan},\ and\
  \citenamefont {Hagan}}]{Chelakkot2014}%
  \BibitemOpen
  \bibfield  {author} {\bibinfo {author} {\bibfnamefont {R.}~\bibnamefont
  {Chelakkot}}, \bibinfo {author} {\bibfnamefont {A.}~\bibnamefont {Gopinath}},
  \bibinfo {author} {\bibfnamefont {L.}~\bibnamefont {Mahadevan}}, \ and\
  \bibinfo {author} {\bibfnamefont {M.~F.}\ \bibnamefont {Hagan}},\ }\href@noop
  {} {\bibfield  {journal} {\bibinfo  {journal} {J. R. Soc. Interface}\
  }\textbf {\bibinfo {volume} {11}} (\bibinfo {year} {2014})}\BibitemShut
  {NoStop}%
\bibitem [{\citenamefont {Wagner}\ \emph {et~al.}(2017)\citenamefont {Wagner},
  \citenamefont {Hagan},\ and\ \citenamefont {Baskaran}}]{wagner2017steady}%
  \BibitemOpen
  \bibfield  {author} {\bibinfo {author} {\bibfnamefont {C.~G.}\ \bibnamefont
  {Wagner}}, \bibinfo {author} {\bibfnamefont {M.~F.}\ \bibnamefont {Hagan}}, \
  and\ \bibinfo {author} {\bibfnamefont {A.}~\bibnamefont {Baskaran}},\
  }\href@noop {} {\bibfield  {journal} {\bibinfo  {journal} {J. Stat. Mech.}\
  }\textbf {\bibinfo {volume} {2017}},\ \bibinfo {pages} {43203} (\bibinfo
  {year} {2017})}\BibitemShut {NoStop}%
\bibitem [{\citenamefont {Weeks}\ \emph {et~al.}(1971)\citenamefont {Weeks},
  \citenamefont {Chandler},\ and\ \citenamefont {Andersen}}]{Weeks1971}%
  \BibitemOpen
  \bibfield  {author} {\bibinfo {author} {\bibfnamefont {J.~D.}\ \bibnamefont
  {Weeks}}, \bibinfo {author} {\bibfnamefont {D.}~\bibnamefont {Chandler}}, \
  and\ \bibinfo {author} {\bibfnamefont {H.~C.}\ \bibnamefont {Andersen}},\
  }\href@noop {} {\bibfield  {journal} {\bibinfo  {journal} {J. Chem. Phys.}\
  }\textbf {\bibinfo {volume} {54}},\ \bibinfo {pages} {5237} (\bibinfo {year}
  {1971})}\BibitemShut {NoStop}%
\bibitem [{\citenamefont {Bra{\'{n}}ka}\ and\ \citenamefont
  {Heyes}(1999)}]{Branka1999}%
  \BibitemOpen
  \bibfield  {author} {\bibinfo {author} {\bibfnamefont {A.~C.}\ \bibnamefont
  {Bra{\'{n}}ka}}\ and\ \bibinfo {author} {\bibfnamefont {D.~M.}\ \bibnamefont
  {Heyes}},\ }\href@noop {} {\bibfield  {journal} {\bibinfo  {journal} {Phys.
  Rev. E}\ }\textbf {\bibinfo {volume} {60}},\ \bibinfo {pages} {2381}
  (\bibinfo {year} {1999})}\BibitemShut {NoStop}%
\bibitem [{\citenamefont {Chetrite}\ and\ \citenamefont
  {Touchette}(2013)}]{Chetrite2013}%
  \BibitemOpen
  \bibfield  {author} {\bibinfo {author} {\bibfnamefont {R.}~\bibnamefont
  {Chetrite}}\ and\ \bibinfo {author} {\bibfnamefont {H.}~\bibnamefont
  {Touchette}},\ }\href@noop {} {\bibfield  {journal} {\bibinfo  {journal}
  {Phys. Rev. Lett.}\ }\textbf {\bibinfo {volume} {111}},\ \bibinfo {pages} {1}
  (\bibinfo {year} {2013})}\BibitemShut {NoStop}%
\bibitem [{\citenamefont {Garrahan}\ and\ \citenamefont
  {Lesanovsky}(2010)}]{Garrahan2010}%
  \BibitemOpen
  \bibfield  {author} {\bibinfo {author} {\bibfnamefont {J.~P.}\ \bibnamefont
  {Garrahan}}\ and\ \bibinfo {author} {\bibfnamefont {I.}~\bibnamefont
  {Lesanovsky}},\ }\href@noop {} {\bibfield  {journal} {\bibinfo  {journal}
  {Phys. Rev. Lett.}\ }\textbf {\bibinfo {volume} {104}},\ \bibinfo {pages} {1}
  (\bibinfo {year} {2010})}\BibitemShut {NoStop}%
\bibitem [{\citenamefont {Hansen}\ and\ \citenamefont
  {McDonald}(1977)}]{Hansen1977}%
  \BibitemOpen
  \bibfield  {author} {\bibinfo {author} {\bibfnamefont {J.~P.}\ \bibnamefont
  {Hansen}}\ and\ \bibinfo {author} {\bibfnamefont {I.~R.}\ \bibnamefont
  {McDonald}},\ }\href@noop {} {\bibfield  {journal} {\bibinfo  {journal}
  {Elsevier}\ } (\bibinfo {year} {1977})}\BibitemShut {NoStop}%
\bibitem [{\citenamefont {Appert-Rolland}\ \emph {et~al.}(2008)\citenamefont
  {Appert-Rolland}, \citenamefont {Derrida}, \citenamefont {Lecomte},\ and\
  \citenamefont {Van~Wijland}}]{Appert-Rolland2008}%
  \BibitemOpen
  \bibfield  {author} {\bibinfo {author} {\bibfnamefont {C.}~\bibnamefont
  {Appert-Rolland}}, \bibinfo {author} {\bibfnamefont {B.}~\bibnamefont
  {Derrida}}, \bibinfo {author} {\bibfnamefont {V.}~\bibnamefont {Lecomte}}, \
  and\ \bibinfo {author} {\bibfnamefont {F.}~\bibnamefont {Van~Wijland}},\
  }\href@noop {} {\bibfield  {journal} {\bibinfo  {journal} {Phys. Rev. E}\
  }\textbf {\bibinfo {volume} {78}},\ \bibinfo {pages} {1} (\bibinfo {year}
  {2008})}\BibitemShut {NoStop}%
\bibitem [{\citenamefont {Hedges}\ \emph {et~al.}(2009)\citenamefont {Hedges},
  \citenamefont {Jack}, \citenamefont {Garrahan},\ and\ \citenamefont
  {Chandler}}]{Hedges2009}%
  \BibitemOpen
  \bibfield  {author} {\bibinfo {author} {\bibfnamefont {L.~O.}\ \bibnamefont
  {Hedges}}, \bibinfo {author} {\bibfnamefont {R.~L.}\ \bibnamefont {Jack}},
  \bibinfo {author} {\bibfnamefont {J.~P.}\ \bibnamefont {Garrahan}}, \ and\
  \bibinfo {author} {\bibfnamefont {D.}~\bibnamefont {Chandler}},\ }\href@noop
  {} {\ \textbf {\bibinfo {volume} {323}},\ \bibinfo {pages} {1309} (\bibinfo
  {year} {2009})}\BibitemShut {NoStop}%
\bibitem [{\citenamefont {Speck}\ \emph {et~al.}(2015)\citenamefont {Speck},
  \citenamefont {Menzel}, \citenamefont {Bialk{\'{e}}},\ and\ \citenamefont
  {L{\"{o}}wen}}]{Speck2015}%
  \BibitemOpen
  \bibfield  {author} {\bibinfo {author} {\bibfnamefont {T.}~\bibnamefont
  {Speck}}, \bibinfo {author} {\bibfnamefont {A.~M.}\ \bibnamefont {Menzel}},
  \bibinfo {author} {\bibfnamefont {J.}~\bibnamefont {Bialk{\'{e}}}}, \ and\
  \bibinfo {author} {\bibfnamefont {H.}~\bibnamefont {L{\"{o}}wen}},\
  }\href@noop {} {\bibfield  {journal} {\bibinfo  {journal} {J. Chem. Phys.}\
  }\textbf {\bibinfo {volume} {142}} (\bibinfo {year} {2015})}\BibitemShut
  {NoStop}%
\bibitem [{\citenamefont {Bialk{\'{e}}}\ \emph {et~al.}(2013)\citenamefont
  {Bialk{\'{e}}}, \citenamefont {L{\"{o}}wen},\ and\ \citenamefont
  {Speck}}]{Bialke2013}%
  \BibitemOpen
  \bibfield  {author} {\bibinfo {author} {\bibfnamefont {J.}~\bibnamefont
  {Bialk{\'{e}}}}, \bibinfo {author} {\bibfnamefont {H.}~\bibnamefont
  {L{\"{o}}wen}}, \ and\ \bibinfo {author} {\bibfnamefont {T.}~\bibnamefont
  {Speck}},\ }\href@noop {} {\bibfield  {journal} {\bibinfo  {journal}
  {EuroPhys. Lett.}\ }\textbf {\bibinfo {volume} {103}} (\bibinfo {year}
  {2013})}\BibitemShut {NoStop}%
\bibitem [{\citenamefont {Hancock}\ and\ \citenamefont
  {Baskaran}(2017)}]{Hancock2017}%
  \BibitemOpen
  \bibfield  {author} {\bibinfo {author} {\bibfnamefont {B.}~\bibnamefont
  {Hancock}}\ and\ \bibinfo {author} {\bibfnamefont {A.}~\bibnamefont
  {Baskaran}},\ }\href@noop {} {\bibfield  {journal} {\bibinfo  {journal} {J.
  Stat. Mech.}\ }\textbf {\bibinfo {volume} {2017}} (\bibinfo {year}
  {2017})}\BibitemShut {NoStop}%
\bibitem [{\citenamefont {Wittkowski}\ \emph {et~al.}(2017)\citenamefont
  {Wittkowski}, \citenamefont {Stenhammar},\ and\ \citenamefont
  {Cates}}]{Wittkowski2017}%
  \BibitemOpen
  \bibfield  {author} {\bibinfo {author} {\bibfnamefont {R.}~\bibnamefont
  {Wittkowski}}, \bibinfo {author} {\bibfnamefont {J.}~\bibnamefont
  {Stenhammar}}, \ and\ \bibinfo {author} {\bibfnamefont {M.~E.}\ \bibnamefont
  {Cates}},\ }\href@noop {} {\bibfield  {journal} {\bibinfo  {journal} {Phys.
  Rev. E}\ ,\ \bibinfo {pages} {1}} (\bibinfo {year} {2017})}\BibitemShut
  {NoStop}%
\bibitem [{SI()}]{SI}%
  \BibitemOpen
  \href@noop {} {\bibinfo  {journal} {Supplementary information}\ }\BibitemShut
  {NoStop}%
\bibitem [{\citenamefont {Abramowitz}\ \emph {et~al.}(1965)\citenamefont
  {Abramowitz}, \citenamefont {Stegun},\ and\ \citenamefont
  {Miller}}]{Abramowitz1965}%
  \BibitemOpen
\bibfield  {journal} {  }\bibfield  {author} {\bibinfo {author} {\bibfnamefont
  {M.}~\bibnamefont {Abramowitz}}, \bibinfo {author} {\bibfnamefont {I.~A.}\
  \bibnamefont {Stegun}}, \ and\ \bibinfo {author} {\bibfnamefont
  {D.}~\bibnamefont {Miller}},\ }\href
  {http://appliedmechanics.asmedigitalcollection.asme.org/article.aspx?articleid=1396937}
  {\bibfield  {journal} {\bibinfo  {journal} {J. App. Mech.}\ }\textbf
  {\bibinfo {volume} {32}},\ \bibinfo {pages} {239} (\bibinfo {year}
  {1965})}\BibitemShut {NoStop}%
\bibitem [{\citenamefont {Pietzonka}\ \emph {et~al.}(2016)\citenamefont
  {Pietzonka}, \citenamefont {Kleinbeck},\ and\ \citenamefont
  {Seifert}}]{Pietzonka2016}%
  \BibitemOpen
  \bibfield  {author} {\bibinfo {author} {\bibfnamefont {P.}~\bibnamefont
  {Pietzonka}}, \bibinfo {author} {\bibfnamefont {K.}~\bibnamefont
  {Kleinbeck}}, \ and\ \bibinfo {author} {\bibfnamefont {U.}~\bibnamefont
  {Seifert}},\ }\href@noop {} {\bibfield  {journal} {\bibinfo  {journal} {New
  J. Phys.}\ }\textbf {\bibinfo {volume} {18}} (\bibinfo {year}
  {2016})}\BibitemShut {NoStop}%
\bibitem [{\citenamefont {Lecomte}\ and\ \citenamefont
  {Tailleur}(2007)}]{Lecomte2007}%
  \BibitemOpen
  \bibfield  {author} {\bibinfo {author} {\bibfnamefont {V.}~\bibnamefont
  {Lecomte}}\ and\ \bibinfo {author} {\bibfnamefont {J.}~\bibnamefont
  {Tailleur}},\ }\href@noop {} {\bibfield  {journal} {\bibinfo  {journal} {J.
  Stat. Mech.}\ } (\bibinfo {year} {2007})}\BibitemShut {NoStop}%
\bibitem [{\citenamefont {Giardin{\`{a}}}\ \emph {et~al.}(2006)\citenamefont
  {Giardin{\`{a}}}, \citenamefont {Kurchan},\ and\ \citenamefont
  {Peliti}}]{Giardina2006}%
  \BibitemOpen
  \bibfield  {author} {\bibinfo {author} {\bibfnamefont {C.}~\bibnamefont
  {Giardin{\`{a}}}}, \bibinfo {author} {\bibfnamefont {J.}~\bibnamefont
  {Kurchan}}, \ and\ \bibinfo {author} {\bibfnamefont {L.}~\bibnamefont
  {Peliti}},\ }\href@noop {} {\bibfield  {journal} {\bibinfo  {journal} {PRL}\
  }\textbf {\bibinfo {volume} {96}},\ \bibinfo {pages} {1} (\bibinfo {year}
  {2006})}\BibitemShut {NoStop}%
\bibitem [{\citenamefont {Ray}\ \emph {et~al.}(2018)\citenamefont {Ray},
  \citenamefont {Chan},\ and\ \citenamefont {Limmer}}]{ray2017exact}%
  \BibitemOpen
  \bibfield  {author} {\bibinfo {author} {\bibfnamefont {U.}~\bibnamefont
  {Ray}}, \bibinfo {author} {\bibfnamefont {G.~K.}\ \bibnamefont {Chan}}, \
  and\ \bibinfo {author} {\bibfnamefont {D.~T.}\ \bibnamefont {Limmer}},\
  }\href@noop {} {\bibfield  {journal} {\bibinfo  {journal} {Phys. Rev. Lett.}\
  } (\bibinfo {year} {2018})}\BibitemShut {NoStop}%
\bibitem [{\citenamefont {Torquato}\ and\ \citenamefont
  {Stillinger}(2003)}]{Torquato2003}%
  \BibitemOpen
  \bibfield  {author} {\bibinfo {author} {\bibfnamefont {S.}~\bibnamefont
  {Torquato}}\ and\ \bibinfo {author} {\bibfnamefont {F.~H.}\ \bibnamefont
  {Stillinger}},\ }\href@noop {} {\bibfield  {journal} {\bibinfo  {journal}
  {Phys. Rev. E}\ }\textbf {\bibinfo {volume} {68}},\ \bibinfo {pages} {1}
  (\bibinfo {year} {2003})}\BibitemShut {NoStop}%
\bibitem [{\citenamefont {Jack}\ \emph {et~al.}(2015)\citenamefont {Jack},
  \citenamefont {Thompson},\ and\ \citenamefont {Sollich}}]{Jack2015}%
  \BibitemOpen
  \bibfield  {author} {\bibinfo {author} {\bibfnamefont {R.~L.}\ \bibnamefont
  {Jack}}, \bibinfo {author} {\bibfnamefont {I.~R.}\ \bibnamefont {Thompson}},
  \ and\ \bibinfo {author} {\bibfnamefont {P.}~\bibnamefont {Sollich}},\ }\href
  {\doibase 10.1103/PhysRevLett.114.060601} {\bibfield  {journal} {\bibinfo
  {journal} {Physical Review Letters}\ }\textbf {\bibinfo {volume} {114}},\
  \bibinfo {pages} {1} (\bibinfo {year} {2015})}\BibitemShut {NoStop}%
\bibitem [{\citenamefont {Steffenoni}\ \emph {et~al.}(2017)\citenamefont
  {Steffenoni}, \citenamefont {Falasco},\ and\ \citenamefont
  {Kroy}}]{Steffenoni2017}%
  \BibitemOpen
  \bibfield  {author} {\bibinfo {author} {\bibfnamefont {S.}~\bibnamefont
  {Steffenoni}}, \bibinfo {author} {\bibfnamefont {G.}~\bibnamefont {Falasco}},
  \ and\ \bibinfo {author} {\bibfnamefont {K.}~\bibnamefont {Kroy}},\ }\href
  {http://link.aps.org/doi/10.1103/PhysRevE.95.052142} {\bibfield  {journal}
  {\bibinfo  {journal} {Phys. Rev. E}\ }\textbf {\bibinfo {volume} {95}},\
  \bibinfo {pages} {052142} (\bibinfo {year} {2017})}\BibitemShut {NoStop}%
\bibitem [{\citenamefont {Solon}\ \emph
  {et~al.}(2015{\natexlab{a}})\citenamefont {Solon}, \citenamefont
  {Stenhammar}, \citenamefont {Wittkowski}, \citenamefont {Kardar},
  \citenamefont {Kafri}, \citenamefont {Cates},\ and\ \citenamefont
  {Tailleur}}]{Solon2015}%
  \BibitemOpen
  \bibfield  {author} {\bibinfo {author} {\bibfnamefont {A.~P.}\ \bibnamefont
  {Solon}}, \bibinfo {author} {\bibfnamefont {J.}~\bibnamefont {Stenhammar}},
  \bibinfo {author} {\bibfnamefont {R.}~\bibnamefont {Wittkowski}}, \bibinfo
  {author} {\bibfnamefont {M.}~\bibnamefont {Kardar}}, \bibinfo {author}
  {\bibfnamefont {Y.}~\bibnamefont {Kafri}}, \bibinfo {author} {\bibfnamefont
  {M.~E.}\ \bibnamefont {Cates}}, \ and\ \bibinfo {author} {\bibfnamefont
  {J.}~\bibnamefont {Tailleur}},\ }\href@noop {} {\bibfield  {journal}
  {\bibinfo  {journal} {Phys. Rev. Lett.}\ }\textbf {\bibinfo {volume} {114}},\
  \bibinfo {pages} {1} (\bibinfo {year} {2015}{\natexlab{a}})}\BibitemShut
  {NoStop}%
\bibitem [{\citenamefont {Takatori}\ and\ \citenamefont
  {Brady}(2015)}]{Takatori2015}%
  \BibitemOpen
  \bibfield  {author} {\bibinfo {author} {\bibfnamefont {S.~C.}\ \bibnamefont
  {Takatori}}\ and\ \bibinfo {author} {\bibfnamefont {J.~F.}\ \bibnamefont
  {Brady}},\ }\href {http://xlink.rsc.org/?DOI=C5SM01792K} {\bibfield
  {journal} {\bibinfo  {journal} {Soft Matter}\ }\textbf {\bibinfo {volume}
  {11}},\ \bibinfo {pages} {7920} (\bibinfo {year} {2015})}\BibitemShut
  {NoStop}%
\bibitem [{\citenamefont {Palacci}\ \emph {et~al.}(2010)\citenamefont
  {Palacci}, \citenamefont {Cottin-Bizonne}, \citenamefont {Ybert},\ and\
  \citenamefont {Bocquet}}]{Palacci2010}%
  \BibitemOpen
  \bibfield  {author} {\bibinfo {author} {\bibfnamefont {J.}~\bibnamefont
  {Palacci}}, \bibinfo {author} {\bibfnamefont {C.}~\bibnamefont
  {Cottin-Bizonne}}, \bibinfo {author} {\bibfnamefont {C.}~\bibnamefont
  {Ybert}}, \ and\ \bibinfo {author} {\bibfnamefont {L.}~\bibnamefont
  {Bocquet}},\ }\href@noop {} {\bibfield  {journal} {\bibinfo  {journal} {Phys.
  Rev. Lett.}\ }\textbf {\bibinfo {volume} {105}},\ \bibinfo {pages} {1}
  (\bibinfo {year} {2010})}\BibitemShut {NoStop}%
\bibitem [{\citenamefont {Wang}\ \emph {et~al.}(2012)\citenamefont {Wang},
  \citenamefont {Kuo}, \citenamefont {Bae},\ and\ \citenamefont
  {Granick}}]{Wang2012}%
  \BibitemOpen
  \bibfield  {author} {\bibinfo {author} {\bibfnamefont {B.}~\bibnamefont
  {Wang}}, \bibinfo {author} {\bibfnamefont {J.}~\bibnamefont {Kuo}}, \bibinfo
  {author} {\bibfnamefont {S.~C.}\ \bibnamefont {Bae}}, \ and\ \bibinfo
  {author} {\bibfnamefont {S.}~\bibnamefont {Granick}},\ }\href
  {http://dx.doi.org/10.1038/nmat3308} {\bibfield  {journal} {\bibinfo
  {journal} {Nature Materials}\ }\textbf {\bibinfo {volume} {11}},\ \bibinfo
  {pages} {481} (\bibinfo {year} {2012})}\BibitemShut {NoStop}%
\bibitem [{\citenamefont {Stuhrmann}\ \emph {et~al.}(2012)\citenamefont
  {Stuhrmann}, \citenamefont {Soares E~Silva}, \citenamefont {Depken},
  \citenamefont {MacKintosh},\ and\ \citenamefont
  {Koenderink}}]{Stuhrmann2012}%
  \BibitemOpen
  \bibfield  {author} {\bibinfo {author} {\bibfnamefont {B.}~\bibnamefont
  {Stuhrmann}}, \bibinfo {author} {\bibfnamefont {M.}~\bibnamefont {Soares
  E~Silva}}, \bibinfo {author} {\bibfnamefont {M.}~\bibnamefont {Depken}},
  \bibinfo {author} {\bibfnamefont {F.~C.}\ \bibnamefont {MacKintosh}}, \ and\
  \bibinfo {author} {\bibfnamefont {G.~H.}\ \bibnamefont {Koenderink}},\
  }\href@noop {} {\bibfield  {journal} {\bibinfo  {journal} {Phys. Rev. E}\
  }\textbf {\bibinfo {volume} {86}},\ \bibinfo {pages} {1} (\bibinfo {year}
  {2012})}\BibitemShut {NoStop}%
\bibitem [{\citenamefont {Stenhammar}\ \emph {et~al.}(2014)\citenamefont
  {Stenhammar}, \citenamefont {Marenduzzo}, \citenamefont {Allen},\ and\
  \citenamefont {Cates}}]{Stenhammar2014}%
  \BibitemOpen
  \bibfield  {author} {\bibinfo {author} {\bibfnamefont {J.}~\bibnamefont
  {Stenhammar}}, \bibinfo {author} {\bibfnamefont {D.}~\bibnamefont
  {Marenduzzo}}, \bibinfo {author} {\bibfnamefont {R.~J.}\ \bibnamefont
  {Allen}}, \ and\ \bibinfo {author} {\bibfnamefont {M.~E.}\ \bibnamefont
  {Cates}},\ }\href {http://xlink.rsc.org/?DOI=C3SM52813H} {\bibfield
  {journal} {\bibinfo  {journal} {Soft Matter}\ }\textbf {\bibinfo {volume}
  {10}},\ \bibinfo {pages} {1489} (\bibinfo {year} {2014})}\BibitemShut
  {NoStop}%
\bibitem [{\citenamefont {Chakraborti}\ \emph {et~al.}(2016)\citenamefont
  {Chakraborti}, \citenamefont {Mishra},\ and\ \citenamefont
  {Pradhan}}]{Chakraborti2016}%
  \BibitemOpen
  \bibfield  {author} {\bibinfo {author} {\bibfnamefont {S.}~\bibnamefont
  {Chakraborti}}, \bibinfo {author} {\bibfnamefont {S.}~\bibnamefont {Mishra}},
  \ and\ \bibinfo {author} {\bibfnamefont {P.}~\bibnamefont {Pradhan}},\
  }\href@noop {} {\bibfield  {journal} {\bibinfo  {journal} {Phys. Rev. E}\
  }\textbf {\bibinfo {volume} {93}},\ \bibinfo {pages} {1} (\bibinfo {year}
  {2016})}\BibitemShut {NoStop}%
\bibitem [{\citenamefont {Fily}\ and\ \citenamefont
  {Marchetti}(2012)}]{Fily2012}%
  \BibitemOpen
  \bibfield  {author} {\bibinfo {author} {\bibfnamefont {Y.}~\bibnamefont
  {Fily}}\ and\ \bibinfo {author} {\bibfnamefont {M.~C.}\ \bibnamefont
  {Marchetti}},\ }\href@noop {} {\bibfield  {journal} {\bibinfo  {journal}
  {Phys. Rev. Lett.}\ }\textbf {\bibinfo {volume} {108}},\ \bibinfo {pages} {1}
  (\bibinfo {year} {2012})}\BibitemShut {NoStop}%
\bibitem [{\citenamefont {Solon}\ \emph
  {et~al.}(2015{\natexlab{b}})\citenamefont {Solon}, \citenamefont {Cates},\
  and\ \citenamefont {Tailleur}}]{Solon2015a}%
  \BibitemOpen
  \bibfield  {author} {\bibinfo {author} {\bibfnamefont {A.~P.}\ \bibnamefont
  {Solon}}, \bibinfo {author} {\bibfnamefont {M.~E.}\ \bibnamefont {Cates}}, \
  and\ \bibinfo {author} {\bibfnamefont {J.}~\bibnamefont {Tailleur}},\
  }\href@noop {} {\bibfield  {journal} {\bibinfo  {journal} {Euro. Phys. J.}\
  }\textbf {\bibinfo {volume} {224}},\ \bibinfo {pages} {1231} (\bibinfo {year}
  {2015}{\natexlab{b}})}\BibitemShut {NoStop}%
\bibitem [{\citenamefont {Risken}\ and\ \citenamefont
  {Frank}(1996)}]{Risken1996}%
  \BibitemOpen
  \bibfield  {author} {\bibinfo {author} {\bibfnamefont {H.}~\bibnamefont
  {Risken}}\ and\ \bibinfo {author} {\bibfnamefont {T.}~\bibnamefont {Frank}},\
  }\href
  {http://www.amazon.ca/exec/obidos/redirect?tag=citeulike09-20&path=ASIN/354061530X}
  {\bibfield  {journal} {\bibinfo  {journal} {Springer}\ } (\bibinfo {year}
  {1996})}\BibitemShut {NoStop}%
\bibitem [{\citenamefont {Li}\ and\ \citenamefont {Tang}(2009)}]{Li2009}%
  \BibitemOpen
  \bibfield  {author} {\bibinfo {author} {\bibfnamefont {G.}~\bibnamefont
  {Li}}\ and\ \bibinfo {author} {\bibfnamefont {J.~X.}\ \bibnamefont {Tang}},\
  }\href@noop {} {\bibfield  {journal} {\bibinfo  {journal} {Phys. Rev. Lett.}\
  }\textbf {\bibinfo {volume} {103}},\ \bibinfo {pages} {1} (\bibinfo {year}
  {2009})}\BibitemShut {NoStop}%
\bibitem [{\citenamefont {Elgeti}\ and\ \citenamefont
  {Gompper}(2013)}]{Elgeti2013}%
  \BibitemOpen
  \bibfield  {author} {\bibinfo {author} {\bibfnamefont {J.}~\bibnamefont
  {Elgeti}}\ and\ \bibinfo {author} {\bibfnamefont {G.}~\bibnamefont
  {Gompper}},\ }\href@noop {} {\bibfield  {journal} {\bibinfo  {journal} {Euro
  Phys. Lett.}\ }\textbf {\bibinfo {volume} {101}} (\bibinfo {year}
  {2013})}\BibitemShut {NoStop}%
\bibitem [{\citenamefont {Fily}\ \emph {et~al.}(2014)\citenamefont {Fily},
  \citenamefont {Henkes},\ and\ \citenamefont {Marchetti}}]{Fily2014a}%
  \BibitemOpen
  \bibfield  {author} {\bibinfo {author} {\bibfnamefont {Y.}~\bibnamefont
  {Fily}}, \bibinfo {author} {\bibfnamefont {S.}~\bibnamefont {Henkes}}, \ and\
  \bibinfo {author} {\bibfnamefont {M.~C.}\ \bibnamefont {Marchetti}},\
  }\href@noop {} {\bibfield  {journal} {\bibinfo  {journal} {Soft Matter}\
  }\textbf {\bibinfo {volume} {10}},\ \bibinfo {pages} {2132} (\bibinfo {year}
  {2014})}\BibitemShut {NoStop}%
\bibitem [{\citenamefont {Yang}\ \emph {et~al.}(2014)\citenamefont {Yang},
  \citenamefont {Manning},\ and\ \citenamefont {Marchetti}}]{Yang2014}%
  \BibitemOpen
  \bibfield  {author} {\bibinfo {author} {\bibfnamefont {X.}~\bibnamefont
  {Yang}}, \bibinfo {author} {\bibfnamefont {M.~L.}\ \bibnamefont {Manning}}, \
  and\ \bibinfo {author} {\bibfnamefont {M.~C.}\ \bibnamefont {Marchetti}},\
  }\href@noop {} {\bibfield  {journal} {\bibinfo  {journal} {Soft Matter}\
  }\textbf {\bibinfo {volume} {10}},\ \bibinfo {pages} {6477} (\bibinfo {year}
  {2014})}\BibitemShut {NoStop}%
\bibitem [{\citenamefont {Elgeti}\ and\ \citenamefont
  {Gompper}(2015)}]{Elgeti2015a}%
  \BibitemOpen
  \bibfield  {author} {\bibinfo {author} {\bibfnamefont {J.}~\bibnamefont
  {Elgeti}}\ and\ \bibinfo {author} {\bibfnamefont {G.}~\bibnamefont
  {Gompper}},\ }\href@noop {} {\bibfield  {journal} {\bibinfo  {journal} {Euro
  Phys. Lett.}\ }\textbf {\bibinfo {volume} {109}} (\bibinfo {year}
  {2015})}\BibitemShut {NoStop}%
\bibitem [{\citenamefont {Zheng}\ \emph {et~al.}(2013)\citenamefont {Zheng},
  \citenamefont {Ten~Hagen}, \citenamefont {Kaiser}, \citenamefont {Wu},
  \citenamefont {Cui}, \citenamefont {Silber-Li},\ and\ \citenamefont
  {L{\"{o}}wen}}]{Zheng2013}%
  \BibitemOpen
  \bibfield  {author} {\bibinfo {author} {\bibfnamefont {X.}~\bibnamefont
  {Zheng}}, \bibinfo {author} {\bibfnamefont {B.}~\bibnamefont {Ten~Hagen}},
  \bibinfo {author} {\bibfnamefont {A.}~\bibnamefont {Kaiser}}, \bibinfo
  {author} {\bibfnamefont {M.}~\bibnamefont {Wu}}, \bibinfo {author}
  {\bibfnamefont {H.}~\bibnamefont {Cui}}, \bibinfo {author} {\bibfnamefont
  {Z.}~\bibnamefont {Silber-Li}}, \ and\ \bibinfo {author} {\bibfnamefont
  {H.}~\bibnamefont {L{\"{o}}wen}},\ }\href@noop {} {\bibfield  {journal}
  {\bibinfo  {journal} {Phys. Rev. E}\ }\textbf {\bibinfo {volume} {88}},\
  \bibinfo {pages} {1} (\bibinfo {year} {2013})}\BibitemShut {NoStop}%
\bibitem [{\citenamefont {Io}\ \emph {et~al.}(2017)\citenamefont {Io},
  \citenamefont {Chen}, \citenamefont {Yeh},\ and\ \citenamefont
  {Cai}}]{Io2017}%
  \BibitemOpen
  \bibfield  {author} {\bibinfo {author} {\bibfnamefont {C.~W.}\ \bibnamefont
  {Io}}, \bibinfo {author} {\bibfnamefont {T.~Y.}\ \bibnamefont {Chen}},
  \bibinfo {author} {\bibfnamefont {J.~W.}\ \bibnamefont {Yeh}}, \ and\
  \bibinfo {author} {\bibfnamefont {S.~C.}\ \bibnamefont {Cai}},\ }\href@noop
  {} {\bibfield  {journal} {\bibinfo  {journal} {Phys. Rev. E}\ }\textbf
  {\bibinfo {volume} {96}},\ \bibinfo {pages} {1} (\bibinfo {year}
  {2017})}\BibitemShut {NoStop}%
\bibitem [{\citenamefont {Chandler}\ and\ \citenamefont
  {Andersen}(1972)}]{Chandler1972}%
  \BibitemOpen
  \bibfield  {author} {\bibinfo {author} {\bibfnamefont {D.}~\bibnamefont
  {Chandler}}\ and\ \bibinfo {author} {\bibfnamefont {H.~C.}\ \bibnamefont
  {Andersen}},\ }\href@noop {} {\bibfield  {journal} {\bibinfo  {journal} {The
  J. Chem. Phys.}\ }\textbf {\bibinfo {volume} {57}},\ \bibinfo {pages} {1918}
  (\bibinfo {year} {1972})}\BibitemShut {NoStop}%
\bibitem [{\citenamefont {Curro}\ and\ \citenamefont
  {Schweizer}(1987)}]{Curro1987}%
  \BibitemOpen
  \bibfield  {author} {\bibinfo {author} {\bibfnamefont {J.~G.}\ \bibnamefont
  {Curro}}\ and\ \bibinfo {author} {\bibfnamefont {K.~S.}\ \bibnamefont
  {Schweizer}},\ }\href@noop {} {\bibfield  {journal} {\bibinfo  {journal} {The
  J. Chem. Phys.}\ }\textbf {\bibinfo {volume} {87}},\ \bibinfo {pages} {1842}
  (\bibinfo {year} {1987})}\BibitemShut {NoStop}%
\bibitem [{\citenamefont {Whitelam}\ \emph {et~al.}(2018)\citenamefont
  {Whitelam}, \citenamefont {Klymko},\ and\ \citenamefont
  {Mandal}}]{Whitelam2018}%
  \BibitemOpen
  \bibfield  {author} {\bibinfo {author} {\bibfnamefont {S.}~\bibnamefont
  {Whitelam}}, \bibinfo {author} {\bibfnamefont {K.}~\bibnamefont {Klymko}}, \
  and\ \bibinfo {author} {\bibfnamefont {D.}~\bibnamefont {Mandal}},\
  }\href@noop {} {\bibfield  {journal} {\bibinfo  {journal} {The J. Chem.
  Phys.}\ }\textbf {\bibinfo {volume} {148}} (\bibinfo {year}
  {2018})}\BibitemShut {NoStop}%
\bibitem [{\citenamefont {Pitard}\ \emph {et~al.}(2011)\citenamefont {Pitard},
  \citenamefont {Lecomte},\ and\ \citenamefont {Van~Wijland}}]{Pitard2011}%
  \BibitemOpen
  \bibfield  {author} {\bibinfo {author} {\bibfnamefont {E.}~\bibnamefont
  {Pitard}}, \bibinfo {author} {\bibfnamefont {V.}~\bibnamefont {Lecomte}}, \
  and\ \bibinfo {author} {\bibfnamefont {F.}~\bibnamefont {Van~Wijland}},\
  }\href@noop {} {\bibfield  {journal} {\bibinfo  {journal} {Epl}\ }\textbf
  {\bibinfo {volume} {96}} (\bibinfo {year} {2011})}\BibitemShut {NoStop}%
\bibitem [{\citenamefont {Garrahan}\ \emph {et~al.}(2007)\citenamefont
  {Garrahan}, \citenamefont {Jack}, \citenamefont {Lecomte}, \citenamefont
  {Pitard}, \citenamefont {Van~Duijvendijk},\ and\ \citenamefont
  {Van~Wijland}}]{Garrahan2007}%
  \BibitemOpen
  \bibfield  {author} {\bibinfo {author} {\bibfnamefont {J.~P.}\ \bibnamefont
  {Garrahan}}, \bibinfo {author} {\bibfnamefont {R.~L.}\ \bibnamefont {Jack}},
  \bibinfo {author} {\bibfnamefont {V.}~\bibnamefont {Lecomte}}, \bibinfo
  {author} {\bibfnamefont {E.}~\bibnamefont {Pitard}}, \bibinfo {author}
  {\bibfnamefont {K.}~\bibnamefont {Van~Duijvendijk}}, \ and\ \bibinfo {author}
  {\bibfnamefont {F.}~\bibnamefont {Van~Wijland}},\ }\href@noop {} {\bibfield
  {journal} {\bibinfo  {journal} {Phys. Rev. Lett.}\ }\textbf {\bibinfo
  {volume} {98}},\ \bibinfo {pages} {0} (\bibinfo {year} {2007})}\BibitemShut
  {NoStop}%
\bibitem [{\citenamefont {Katira}\ \emph {et~al.}(2017)\citenamefont {Katira},
  \citenamefont {Garrahan},\ and\ \citenamefont {Mandadapu}}]{Katira2017}%
  \BibitemOpen
  \bibfield  {author} {\bibinfo {author} {\bibfnamefont {S.}~\bibnamefont
  {Katira}}, \bibinfo {author} {\bibfnamefont {J.~P.}\ \bibnamefont
  {Garrahan}}, \ and\ \bibinfo {author} {\bibfnamefont {K.~K.}\ \bibnamefont
  {Mandadapu}},\ }\href {http://arxiv.org/abs/1710.04747} {\bibfield  {journal}
  {\bibinfo  {journal} {Phys. Rev. Lett.}\ }\textbf {\bibinfo {volume} {120}},\
  \bibinfo {pages} {260602} (\bibinfo {year} {2017})}\BibitemShut {NoStop}%
\bibitem [{\citenamefont {Mandal}\ \emph {et~al.}(2017)\citenamefont {Mandal},
  \citenamefont {Klymko},\ and\ \citenamefont {DeWeese}}]{Mandal2017}%
  \BibitemOpen
  \bibfield  {author} {\bibinfo {author} {\bibfnamefont {D.}~\bibnamefont
  {Mandal}}, \bibinfo {author} {\bibfnamefont {K.}~\bibnamefont {Klymko}}, \
  and\ \bibinfo {author} {\bibfnamefont {M.~R.}\ \bibnamefont {DeWeese}},\
  }\href@noop {} {\bibfield  {journal} {\bibinfo  {journal} {Phys. Rev. Lett.}\
  }\textbf {\bibinfo {volume} {119}},\ \bibinfo {pages} {1} (\bibinfo {year}
  {2017})}\BibitemShut {NoStop}%
\bibitem [{\citenamefont {Fodor}\ \emph {et~al.}(2016)\citenamefont {Fodor},
  \citenamefont {Nardini}, \citenamefont {Cates}, \citenamefont {Tailleur},
  \citenamefont {Visco},\ and\ \citenamefont {Van~Wijland}}]{Fodor2016}%
  \BibitemOpen
  \bibfield  {author} {\bibinfo {author} {\bibfnamefont {E.}~\bibnamefont
  {Fodor}}, \bibinfo {author} {\bibfnamefont {C.}~\bibnamefont {Nardini}},
  \bibinfo {author} {\bibfnamefont {M.~E.}\ \bibnamefont {Cates}}, \bibinfo
  {author} {\bibfnamefont {J.}~\bibnamefont {Tailleur}}, \bibinfo {author}
  {\bibfnamefont {P.}~\bibnamefont {Visco}}, \ and\ \bibinfo {author}
  {\bibfnamefont {F.}~\bibnamefont {Van~Wijland}},\ }\href@noop {} {\bibfield
  {journal} {\bibinfo  {journal} {Phys. Rev. Lett.}\ }\textbf {\bibinfo
  {volume} {117}},\ \bibinfo {pages} {1} (\bibinfo {year} {2016})}\BibitemShut
  {NoStop}%
\bibitem [{\citenamefont {Mehl}\ \emph {et~al.}(2008)\citenamefont {Mehl},
  \citenamefont {Speck},\ and\ \citenamefont {Seifert}}]{Mehl2008}%
  \BibitemOpen
  \bibfield  {author} {\bibinfo {author} {\bibfnamefont {J.}~\bibnamefont
  {Mehl}}, \bibinfo {author} {\bibfnamefont {T.}~\bibnamefont {Speck}}, \ and\
  \bibinfo {author} {\bibfnamefont {U.}~\bibnamefont {Seifert}},\ }\href@noop
  {} {\bibfield  {journal} {\bibinfo  {journal} {Phys. Rev. E}\ }\textbf
  {\bibinfo {volume} {78}},\ \bibinfo {pages} {1} (\bibinfo {year}
  {2008})}\BibitemShut {NoStop}%
\bibitem [{\citenamefont {Speck}\ \emph {et~al.}(2012)\citenamefont {Speck},
  \citenamefont {Engel},\ and\ \citenamefont {Seifert}}]{Speck2012}%
  \BibitemOpen
  \bibfield  {author} {\bibinfo {author} {\bibfnamefont {T.}~\bibnamefont
  {Speck}}, \bibinfo {author} {\bibfnamefont {A.}~\bibnamefont {Engel}}, \ and\
  \bibinfo {author} {\bibfnamefont {U.}~\bibnamefont {Seifert}},\ }\href@noop
  {} {\bibfield  {journal} {\bibinfo  {journal} {J. Stat. Mech.}\ }\textbf
  {\bibinfo {volume} {2012}} (\bibinfo {year} {2012})}\BibitemShut {NoStop}%
\bibitem [{\citenamefont {Derrida}\ \emph {et~al.}(2001)\citenamefont
  {Derrida}, \citenamefont {Lebowitz},\ and\ \citenamefont
  {Speer}}]{Derrida2001}%
  \BibitemOpen
  \bibfield  {author} {\bibinfo {author} {\bibfnamefont {B.}~\bibnamefont
  {Derrida}}, \bibinfo {author} {\bibfnamefont {J.~L.}\ \bibnamefont
  {Lebowitz}}, \ and\ \bibinfo {author} {\bibfnamefont {E.~R.}\ \bibnamefont
  {Speer}},\ }\href@noop {} {\bibfield  {journal} {\bibinfo  {journal} {PRL}\
  ,\ \bibinfo {pages} {1}} (\bibinfo {year} {2001})}\BibitemShut {NoStop}%
\end{thebibliography}

\begin{thebibliography}{11}%
\makeatletter
\providecommand \@ifxundefined [1]{%
 \@ifx{#1\undefined}
}%
\providecommand \@ifnum [1]{%
 \ifnum #1\expandafter \@firstoftwo
 \else \expandafter \@secondoftwo
 \fi
}%
\providecommand \@ifx [1]{%
 \ifx #1\expandafter \@firstoftwo
 \else \expandafter \@secondoftwo
 \fi
}%
\providecommand \natexlab [1]{#1}%
\providecommand \enquote  [1]{``#1''}%
\providecommand \bibnamefont  [1]{#1}%
\providecommand \bibfnamefont [1]{#1}%
\providecommand \citenamefont [1]{#1}%
\providecommand \href@noop [0]{\@secondoftwo}%
\providecommand \href [0]{\begingroup \@sanitize@url \@href}%
\providecommand \@href[1]{\@@startlink{#1}\@@href}%
\providecommand \@@href[1]{\endgroup#1\@@endlink}%
\providecommand \@sanitize@url [0]{\catcode `\\12\catcode `\$12\catcode
  `\&12\catcode `\#12\catcode `\^12\catcode `\_12\catcode `\%12\relax}%
\providecommand \@@startlink[1]{}%
\providecommand \@@endlink[0]{}%
\providecommand \url  [0]{\begingroup\@sanitize@url \@url }%
\providecommand \@url [1]{\endgroup\@href {#1}{\urlprefix }}%
\providecommand \urlprefix  [0]{URL }%
\providecommand \Eprint [0]{\href }%
\providecommand \doibase [0]{http://dx.doi.org/}%
\providecommand \selectlanguage [0]{\@gobble}%
\providecommand \bibinfo  [0]{\@secondoftwo}%
\providecommand \bibfield  [0]{\@secondoftwo}%
\providecommand \translation [1]{[#1]}%
\providecommand \BibitemOpen [0]{}%
\providecommand \bibitemStop [0]{}%
\providecommand \bibitemNoStop [0]{.\EOS\space}%
\providecommand \EOS [0]{\spacefactor3000\relax}%
\providecommand \BibitemShut  [1]{\csname bibitem#1\endcsname}%
\let\auto@bib@innerbib\@empty
\bibitem [{\citenamefont {Pietzonka}\ \emph {et~al.}(2016)\citenamefont
  {Pietzonka}, \citenamefont {Kleinbeck},\ and\ \citenamefont
  {Seifert}}]{Pietzonka2016}%
  \BibitemOpen
  \bibfield  {author} {\bibinfo {author} {\bibfnamefont {P.}~\bibnamefont
  {Pietzonka}}, \bibinfo {author} {\bibfnamefont {K.}~\bibnamefont
  {Kleinbeck}}, \ and\ \bibinfo {author} {\bibfnamefont {U.}~\bibnamefont
  {Seifert}},\ }\href@noop {} {\bibfield  {journal} {\bibinfo  {journal} {New
  J. Phys.}\ }\textbf {\bibinfo {volume} {18}} (\bibinfo {year}
  {2016})}\BibitemShut {NoStop}%
\bibitem [{\citenamefont {Chetrite}\ and\ \citenamefont
  {Touchette}(2015)}]{Chetrite2015}%
  \BibitemOpen
  \bibfield  {author} {\bibinfo {author} {\bibfnamefont {R.}~\bibnamefont
  {Chetrite}}\ and\ \bibinfo {author} {\bibfnamefont {H.}~\bibnamefont
  {Touchette}},\ }\href@noop {} {\bibfield  {journal} {\bibinfo  {journal}
  {Ann. Hen. P.}\ }\textbf {\bibinfo {volume} {16}},\ \bibinfo {pages} {2005}
  (\bibinfo {year} {2015})}\BibitemShut {NoStop}%
\bibitem [{\citenamefont {Redner}\ \emph {et~al.}(2013)\citenamefont {Redner},
  \citenamefont {Hagan},\ and\ \citenamefont {Baskaran}}]{Redner2013}%
  \BibitemOpen
  \bibfield  {author} {\bibinfo {author} {\bibfnamefont {G.~S.}\ \bibnamefont
  {Redner}}, \bibinfo {author} {\bibfnamefont {M.~F.}\ \bibnamefont {Hagan}}, \
  and\ \bibinfo {author} {\bibfnamefont {A.}~\bibnamefont {Baskaran}},\
  }\href@noop {} {\bibfield  {journal} {\bibinfo  {journal} {Phys. Rev. Lett.}\
  }\textbf {\bibinfo {volume} {110}},\ \bibinfo {pages} {1} (\bibinfo {year}
  {2013})}\BibitemShut {NoStop}%
\bibitem [{\citenamefont {Ray}\ \emph {et~al.}(2018)\citenamefont {Ray},
  \citenamefont {Chan},\ and\ \citenamefont {Limmer}}]{ray2017exact}%
  \BibitemOpen
  \bibfield  {author} {\bibinfo {author} {\bibfnamefont {U.}~\bibnamefont
  {Ray}}, \bibinfo {author} {\bibfnamefont {G.~K.}\ \bibnamefont {Chan}}, \
  and\ \bibinfo {author} {\bibfnamefont {D.~T.}\ \bibnamefont {Limmer}},\
  }\href@noop {} {\bibfield  {journal} {\bibinfo  {journal} {Phys. Rev. Lett.}\
  } (\bibinfo {year} {2018})}\BibitemShut {NoStop}%
\bibitem [{\citenamefont {Cates}\ and\ \citenamefont
  {Tailleur}(2015)}]{Cates2015}%
  \BibitemOpen
  \bibfield  {author} {\bibinfo {author} {\bibfnamefont {M.~E.}\ \bibnamefont
  {Cates}}\ and\ \bibinfo {author} {\bibfnamefont {J.}~\bibnamefont
  {Tailleur}},\ }\href
  {http://www.annualreviews.org/doi/10.1146/annurev-conmatphys-031214-014710}
  {\bibfield  {journal} {\bibinfo  {journal} {Ann. Rev. Cond. Mat. Phys.}\
  }\textbf {\bibinfo {volume} {6}},\ \bibinfo {pages} {219} (\bibinfo {year}
  {2015})}\BibitemShut {NoStop}%
\bibitem [{\citenamefont {Speck}\ \emph {et~al.}(2015)\citenamefont {Speck},
  \citenamefont {Menzel}, \citenamefont {Bialk{\'{e}}},\ and\ \citenamefont
  {L{\"{o}}wen}}]{Speck2015}%
  \BibitemOpen
  \bibfield  {author} {\bibinfo {author} {\bibfnamefont {T.}~\bibnamefont
  {Speck}}, \bibinfo {author} {\bibfnamefont {A.~M.}\ \bibnamefont {Menzel}},
  \bibinfo {author} {\bibfnamefont {J.}~\bibnamefont {Bialk{\'{e}}}}, \ and\
  \bibinfo {author} {\bibfnamefont {H.}~\bibnamefont {L{\"{o}}wen}},\
  }\href@noop {} {\bibfield  {journal} {\bibinfo  {journal} {J. Chem. Phys.}\
  }\textbf {\bibinfo {volume} {142}} (\bibinfo {year} {2015})}\BibitemShut
  {NoStop}%
\bibitem [{\citenamefont {Stenhammar}\ \emph {et~al.}(2014)\citenamefont
  {Stenhammar}, \citenamefont {Marenduzzo}, \citenamefont {Allen},\ and\
  \citenamefont {Cates}}]{Stenhammar2014}%
  \BibitemOpen
  \bibfield  {author} {\bibinfo {author} {\bibfnamefont {J.}~\bibnamefont
  {Stenhammar}}, \bibinfo {author} {\bibfnamefont {D.}~\bibnamefont
  {Marenduzzo}}, \bibinfo {author} {\bibfnamefont {R.~J.}\ \bibnamefont
  {Allen}}, \ and\ \bibinfo {author} {\bibfnamefont {M.~E.}\ \bibnamefont
  {Cates}},\ }\href {http://xlink.rsc.org/?DOI=C3SM52813H} {\bibfield
  {journal} {\bibinfo  {journal} {Soft Matter}\ }\textbf {\bibinfo {volume}
  {10}},\ \bibinfo {pages} {1489} (\bibinfo {year} {2014})}\BibitemShut
  {NoStop}%
\bibitem [{\citenamefont {Fily}\ and\ \citenamefont
  {Marchetti}(2012)}]{Fily2012}%
  \BibitemOpen
  \bibfield  {author} {\bibinfo {author} {\bibfnamefont {Y.}~\bibnamefont
  {Fily}}\ and\ \bibinfo {author} {\bibfnamefont {M.~C.}\ \bibnamefont
  {Marchetti}},\ }\href@noop {} {\bibfield  {journal} {\bibinfo  {journal}
  {Phys. Rev. Lett.}\ }\textbf {\bibinfo {volume} {108}},\ \bibinfo {pages} {1}
  (\bibinfo {year} {2012})}\BibitemShut {NoStop}%
\bibitem [{\citenamefont {Fily}\ \emph {et~al.}(2014)\citenamefont {Fily},
  \citenamefont {Henkes},\ and\ \citenamefont {Marchetti}}]{Fily2014}%
  \BibitemOpen
  \bibfield  {author} {\bibinfo {author} {\bibfnamefont {Y.}~\bibnamefont
  {Fily}}, \bibinfo {author} {\bibfnamefont {S.}~\bibnamefont {Henkes}}, \ and\
  \bibinfo {author} {\bibfnamefont {M.~C.}\ \bibnamefont {Marchetti}},\ }\href
  {http://xlink.rsc.org/?DOI=C3SM52469H} {\bibfield  {journal} {\bibinfo
  {journal} {Soft Matter}\ }\textbf {\bibinfo {volume} {10}},\ \bibinfo {pages}
  {2132} (\bibinfo {year} {2014})}\BibitemShut {NoStop}%
\bibitem [{\citenamefont {Andrieux}\ and\ \citenamefont
  {Gaspard}(2004)}]{Andrieux2004}%
  \BibitemOpen
  \bibfield  {author} {\bibinfo {author} {\bibfnamefont {D.}~\bibnamefont
  {Andrieux}}\ and\ \bibinfo {author} {\bibfnamefont {P.}~\bibnamefont
  {Gaspard}},\ }\href {\doibase 10.1063/1.1782391} {\bibfield  {journal}
  {\bibinfo  {journal} {J. Chem. Phys.}\ }\textbf {\bibinfo {volume} {121}},\
  \bibinfo {pages} {6167} (\bibinfo {year} {2004})}\BibitemShut {NoStop}%
\bibitem [{\citenamefont {Cates}\ and\ \citenamefont
  {Tailleur}(2013)}]{Cates2013}%
  \BibitemOpen
  \bibfield  {author} {\bibinfo {author} {\bibfnamefont {M.~E.}\ \bibnamefont
  {Cates}}\ and\ \bibinfo {author} {\bibfnamefont {J.}~\bibnamefont
  {Tailleur}},\ }\href@noop {} {\bibfield  {journal} {\bibinfo  {journal}
  {Europhys. Lett.}\ }\textbf {\bibinfo {volume} {101}} (\bibinfo {year}
  {2013})}\BibitemShut {NoStop}%
\end{thebibliography}
\end{document}